\documentclass[11pt]{article}

\usepackage{algorithm} 
\usepackage{algpseudocode} 
\usepackage{amsmath}
\usepackage{amsfonts}
\usepackage{amssymb}
\usepackage[english]{babel}
\usepackage{booktabs}
\usepackage{caption}
\usepackage{cite}
\usepackage{doi}
\usepackage{enumitem}
\usepackage{float}
\usepackage{graphicx}
\usepackage{hyperref}
\usepackage[capitalise]{cleveref}
\usepackage[utf8]{inputenc}
\usepackage{mathptmx}
\usepackage{microtype}
\usepackage{siunitx}
\usepackage{stfloats}
\usepackage{subfig}
\usepackage{tabularx}
\usepackage{textcomp}
\usepackage{xspace}
\usepackage{color}

\usepackage{./template/arxiv}

\usepackage{./input/ao-math-std}
\usepackage{./input/ao-math-graphs}
\usepackage{./input/ao-math-fields}
  
\DeclareMathOperator\dist{dist}
\DeclareSIUnit\mmHg{mmHg}

\newcommand\perf{\tsb{perf}}
\newcommand\term{\tsb{term}}
\newcommand\topo{\tsb{topo}}

\newcommand\qfor{\quad\forall}
\newcommand\surface{\boundary\Omega}
\newcommand\extArcs{\Arcs\tsb{ext}}
\newcommand\extNodes{\Nodes\tsb{ext}}
\newcommand\setE{\graphset E}
\newcommand\setB{\graphset B}
\newcommand\setF{\graphset F}
\newcommand\setP{\graphset P}
\newcommand\FLe{Fåhræus--Lindqvist effect\xspace}

\newcommand\cardL[1][]{\card{\Leaves_{#1}}}

\newcommand\orcauth[2]{\href{https://orcid.org/#1}
  {\includegraphics[height=0.7em]{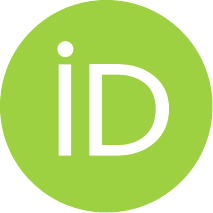}\enspace#2}}
\title{Optimizing Non-Intersecting Synthetic Vascular Trees in Nonconvex Organs}
\date{}

\author{
    \orcauth{0000-0003-0276-6029}{Etienne Jessen}\\
	Institute for Mechanics, Computational Mechanics Group\\
	Technical University of Darmstadt\\
	64287 Darmstadt, Germany\\
	\texttt{etienne.jessen@tu-darmstadt.de} \\
\And
    \orcauth{0000-0002-6343-9809}{Marc C. Steinbach} \\
	Institute of Applied Mathematics\\
	Leibniz Universität Hannover\\
	30167 Hannover, Germany\\
	\texttt{mcs@ifam.uni-hannover.de} \\
\And
    \orcauth{0000-0002-9068-6311}{Dominik Schillinger} \\
	Institute for Mechanics, Computational Mechanics Group\\
	Technical University of Darmstadt\\
	64287 Darmstadt, Germany\\
	\texttt{dominik.schillinger@tu-darmstadt.de} \\
}

\begin{document}
\maketitle

\begin{abstract}
\textit{Objective:}
The understanding of the mechanisms driving vascular development is still limited.
Techniques to generate vascular trees synthetically have been developed to tackle this problem.
However, most algorithms are limited to single trees inside convex perfusion volumes.
We introduce a new framework for generating multiple trees inside general nonconvex perfusion volumes.
\textit{Methods:}
Our framework combines topology optimization and global geometry optimization into a single algorithmic
approach.
Our first contribution is defining a baseline problem based on Murray’s original formulation, which accommodates efficient solution algorithms.
The problem of finding the global minimum is cast into a nonlinear optimization problem (NLP) with merely super-linear solution effort.
Our second contribution extends the NLP to constrain multiple vascular trees inside any nonconvex boundary while avoiding intersections.
We test our framework against a benchmark of an anatomic region of brain tissue and a vasculature of the human liver.
\textit{Results:}
In all cases, the total tree energy is improved significantly compared to local approaches.
\textit{Conclusion:}
By avoiding intersections globally, we can reproduce key physiological features such as parallel running inflow vessels and tortuous vessels.
\textit{Significance:}
The ability to generate non-intersecting vascular trees inside nonconvex organs can improve the functional assessment of organs.
\end{abstract}

\newpage

\newpage
\newpage
\section{Introduction}
\label{sec:introduction3}
The vasculature of the human body is responsible for supplying cells with nutrients by permitting blood to circulate throughout the body \cite{noordergraaf2012circulatory}.
Here, vascular trees are responsible for either distributing (supplying) or collecting (draining) the blood to and from the micro-circulation.
These trees are hierarchical and reach from the largest arteries and veins (\SI{10}{cm}) to the smallest arterioles/venules (\SI{50}{\um}) and are the target of many diseases, e.g., abdominal aneurysms \cite{sakalihasan2005abdominal} or stroke in the brain \cite{bonita1992epidemiolgy}.
To this date, there is only limited understanding of how these diseases develop and how they can be detected early on.
One major reason is that the understanding of what drives the development of vascular trees across the different scales and what constitutes a healthy vascular tree is still limited.
Consequently, a better understanding could elevate the functional assessment
of organs based on available (patient-specific) imaging data.

One promising idea to achieve a better understanding is to generate a vascular tree synthetically.
This was first conceived by Schreiner \cite{schreiner1993computer} in 1993 and was later extended by Karch to three dimensions \cite{karch1999three}.
They used a method called Constrained Constructive Optimization (CCO) to generate trees iteratively by adding a new terminal node at each iteration.
To this day most generation methods are based on CCO, and extensions were made to fit the method to specific organs, e.g., the liver \cite{schwen2012analysis, kretowski2003physiologically, jessen2022rigorous}, brain \cite{sexton2023rapid} or retina \cite{talou2021adaptive}.
Some work also went into improving other aspects of the method, such as including nonconvex regions \cite{cury2021parallel} or multiple trees \cite{guy20193d}.
However, to this day, CCO is still limited to local optimality principles.
Alternative methods such as Global Constructive Optimization (GCO) \cite{hahn2005fractal} and Simulated Annealing \cite{keelan2016simulated} were developed as a response but are still based on local optimizations and suffer from poor computational complexity.
Consequently, a quantitative comparison against real vascular trees remains a challenge.
In \cite{jessen2022rigorous},
we used rigorous mathematical optimization methods to alleviate some of these problems by extending the algorithm to consider the global geometry and topology.
In \cite{jessen2023branching},
we extended the model further to include
a non-Newtonian blood model in the form of the \FLe.
However, during all studies, we only considered a single tree
inside a perfusion domain without severe nonconvex features.

In this paper, we extend the framework introduced
in \cite{jessen2022rigorous,jessen2023branching}
to generate multiple non-intersecting vascular trees inside the same
(generally nonconvex) perfusion volume.
We utilize our global optimization framework to enforce
non-intersection of all trees and containment inside the volume
(semi-) globally.
Lastly, we test our new approach
for a brain region with severe nonconvexity
and for the complete liver with three distinct vascular trees.

\section{Methods}
\subsection{Definitions and assumptions}\label{sec:definitions3}
\begin{figure}[ht]
  \centering
  \includegraphics[width=0.53\textwidth]{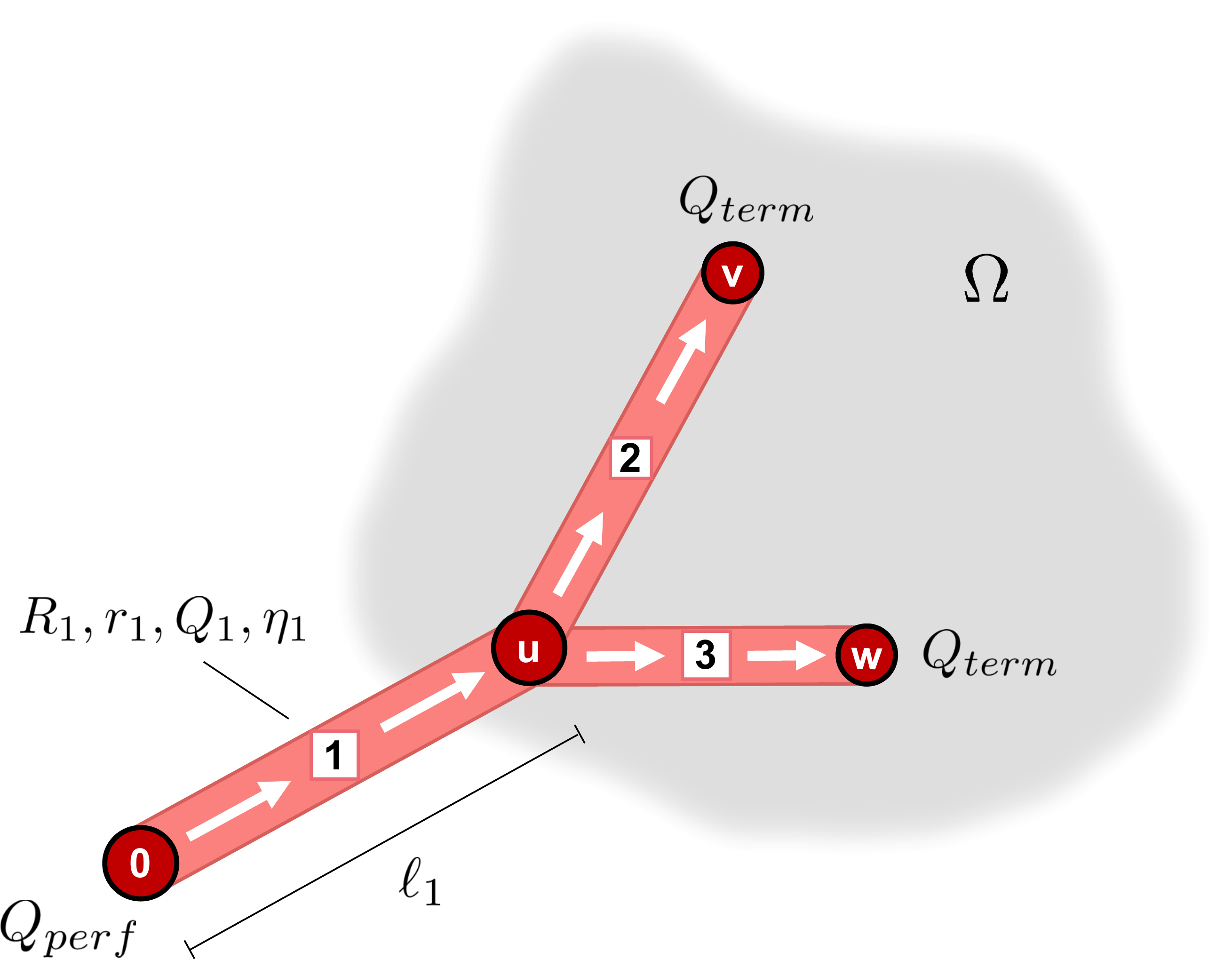}
  \caption{Schematic of a vascular tree and its relation to nodes and segments.
    Red circles denote a node, white rectangles denote a segment.
    This tree has a given inflow $Q_1 = Q\perf$
    and equal terminal outflow $Q_2 = Q_3 = Q\term$
    through each of the outlets (leaves).
    Arrows indicate the flow direction from the root node to the terminal nodes.
    All branching and terminal nodes are
    inside the (nonconvex) perfusion domain~$\Omega$.}
  \label{fig:tree-structure}
\end{figure}

The perfusion domain of an organ of interest is denoted by $\Omega$.
We describe the topology of each vascular tree as a directed graph
$\Tree = (\Nodes, \Arcs)$ with nodes $u \in \Nodes$ and segments $a \in \Arcs$.
The \emph{root} (index $0$) is the proximal node of the single root segment,
and each node distal to a terminal segment is a \emph{leave} $v \in \Leaves$.
The geometry of each tree consists
of the geometric location $x_u$ of each node $u$,
and the length $\ell_a = \norm{x_u - x_v}$, radius $r_a$
and volumetric flow $Q_a$ of each segment $a = uv$.
A segment thus simplifies a \emph{vessel}
to a rigid and straight cylindrical tube,
schematically shown in \cref{fig:tree-structure}.

The blood flow through each vessel is assumed to be laminar and we approximate blood as an incompressible, homogeneous Newtonian fluid.
Poiseuille's law then describes the flow through each segment with
\begin{equation}
  \Delta p_a = R_a Q_a \qfor a \in \Arcs,
\end{equation}
where $\Delta p_a = p_u - p_v$ is the pressure drop across segment $a=uv$.
The hydrodynamic resistance $R_a$ of each segment $a$ follows with
\begin{equation}
  R_a = \frac{8 \eta}{\pi} \frac{\ell_a}{r_a^4} \qfor a \in \Arcs,
\end{equation}
where $\eta$ is the dynamic viscosity of blood.
Each tree is perfused at steady-state by a given perfusion (root) flow $Q\perf$.
We note that any (arbitrary) distribution of flow
at the leaves can be modeled in our framework.
A standard assumption is
a homogeneous distribution of $Q\perf$ to all $N$ leaves,
resulting in a terminal flow $Q\term = Q\perf / N$.
In any case, Kirchhoff's law describes
the flow distributions at the branching nodes with
\begin{equation}\label{eq:conserved-flow}
  Q_{uv} = \sum_{vw \in \Arcs} Q_{vw}
  \qfor v \in \Nodes \setminus (\set{0} \cup \Leaves).
\end{equation}
The trees are assumed to obey scaling relations
based on minimizing certain cost functions.
Murray \cite{murray1926physiological} proposed
that each vessel minimizes its total power,
which consists of the power to maintain blood inside the vessel, $P\tsb{vol}$,
and the (viscous) power to move blood through the vessel, $P\tsb{vis}$.
The resulting scaling relation, known as Murray's law,
describes the relationship between the radius of a parent vessel ($r_0$)
and its children's vessels ($r_1, \dots, r_n$) with
\begin{align}
  r_0^3 = r_1^3 + \dots + r_n^3.
\end{align}
We extend Murray's formulation to the entire tree,
which results in the cost function
\begin{equation} \label{eq:murray-cost}
  f_\Tree
  =
  P\tsb{vol} + P\tsb{vis}
  =
  \sum_{a \in \Arcs}
  m_b \pi \ell_a r_a^2 + \frac{8 \eta}{\pi} \frac{\ell_a}{r_a^4} Q_a^2
  .
\end{equation}
The parameter $m_b$ is the metabolic demand of blood,
measured in \si{\uW\per\cubic\mm}.
It describes the accumulated energy expenditure of blood
(blood plasma and RBCs) and vessels.
More details on this parameter can be found in \cite{liu2007vascular}.

\subsection{Derivation of baseline problem}
From these assumptions we derive an optimization problem
that is used as the basis for our generation framework.
We start by rewriting \eqref{eq:murray-cost} to
\begin{align}
  \label{eq:weighted-length}
  f_\Tree &= \sum_{a \in \Arcs} \ell_a w_a(r_a),
\end{align}
where $w_a$ is the power weight function defined at each segment $a$ with
given parameters $m_b$, $\eta$ and $Q_a$,
\begin{align}
  w_a(r_a) \define
  w(r_a; m_b, \eta, Q_a) \define
  m_b \pi r_a^2 + \frac{8 \eta}{\pi r_a^4} Q_a^2.
\end{align}
Our assumptions do not include any global coupling constraints between nodes.
As a result, the (globally) optimal radius of each segment $a$ can be independently computed by solving
\begin{align}
  \pfrac{w}{r_a} = 2m_b \pi r_a - \frac{32 \eta Q_a^2}{\pi r_a^5} = 0,
\end{align}
leading to
\begin{align}
  \label{eq:radius-murray}
  r_a = \sqrt[6]{\frac{16 \eta}{m_b \pi^2} Q_a^2}.
\end{align}
Since we enforce conservation of mass at each branch with
Kirchhoff's law \eqref{eq:conserved-flow},
we satisfy Murray's law automatically.
This is to be expected, as we derived our solution from his original problem.
We note, though, that the validity of Murray's law
depends on the type of optimization problem considered,
as highlighted in \cite{jessen2023branching}.
By assuming a constant blood viscosity $\eta$ and metabolic demand $m_b$,
we can simplify \eqref{eq:radius-murray} to
\begin{align}
  r_a &= c_1 \sqrt[3]{Q_a} \qfor a \in \Arcs,
  & c_1 &\define \sqrt[6]{\frac{16 \eta}{m_b \pi^2}}.
\end{align}
If we further assume equal outflow, the flow across vessel $a$ simply becomes
\begin{align}
  \label{eq:radius-murray-simplified-1}
  Q_a = Q_{uv} = Q_0 \frac{\cardL[u]}{\cardL[0]}
  \qfor a \in \Arcs,
\end{align}
where $\cardL[u]$ is the number of leave nodes downstream of node $u$
(with $\Leaves_0 = \Leaves$).
In this case, the radius only depends on the topology of the tree with
\begin{align}
  \label{eq:radius-murray-simplified-2}
  r_a &= r_{uv} = c_2 \sqrt[3]{\cardL[u]} \qfor a \in \Arcs,
  & c_2 &= c_1 \sqrt[3]{\frac{Q_0}{\cardL[0]}}.
\end{align}

Both $c_1$ and $c_2$ are constants that can be computed
before starting our solution algorithm,
based on the choice of flow distribution.
The problem of finding the optimal geometry of a tree
then only consists in finding the optimal nodal positions
(and corresponding vessel lengths).

\subsection{Geometry optimization}\label{sec:tree_geometry}
\subsubsection{NLP formulation of global geometry}
Current state-of-the-art generation methods for synthetic vascular trees,
such as \cite{cury2021parallel, sexton2023rapid},
only perform a local optimization
of each single branch point
when ``growing'' the tree node by node.
In \cite{jessen2022rigorous},
we extended this process to cover the global geometry,
which included the positions and pressures at all branching nodes
as well as the lengths and radii of all arcs.
The problem was cast into a nonlinear optimization problem (NLP),
which admits efficient solution algorithms.
Based on this approach,
we now construct an NLP for our baseline problem.
We note that various other design goals and constraints
can also be modeled simply by
adapting or extending our baseline problem.

\begin{figure}[ht]
  \centering
  \includegraphics[trim=0 0 0 20, clip=true, width=0.49\textwidth]{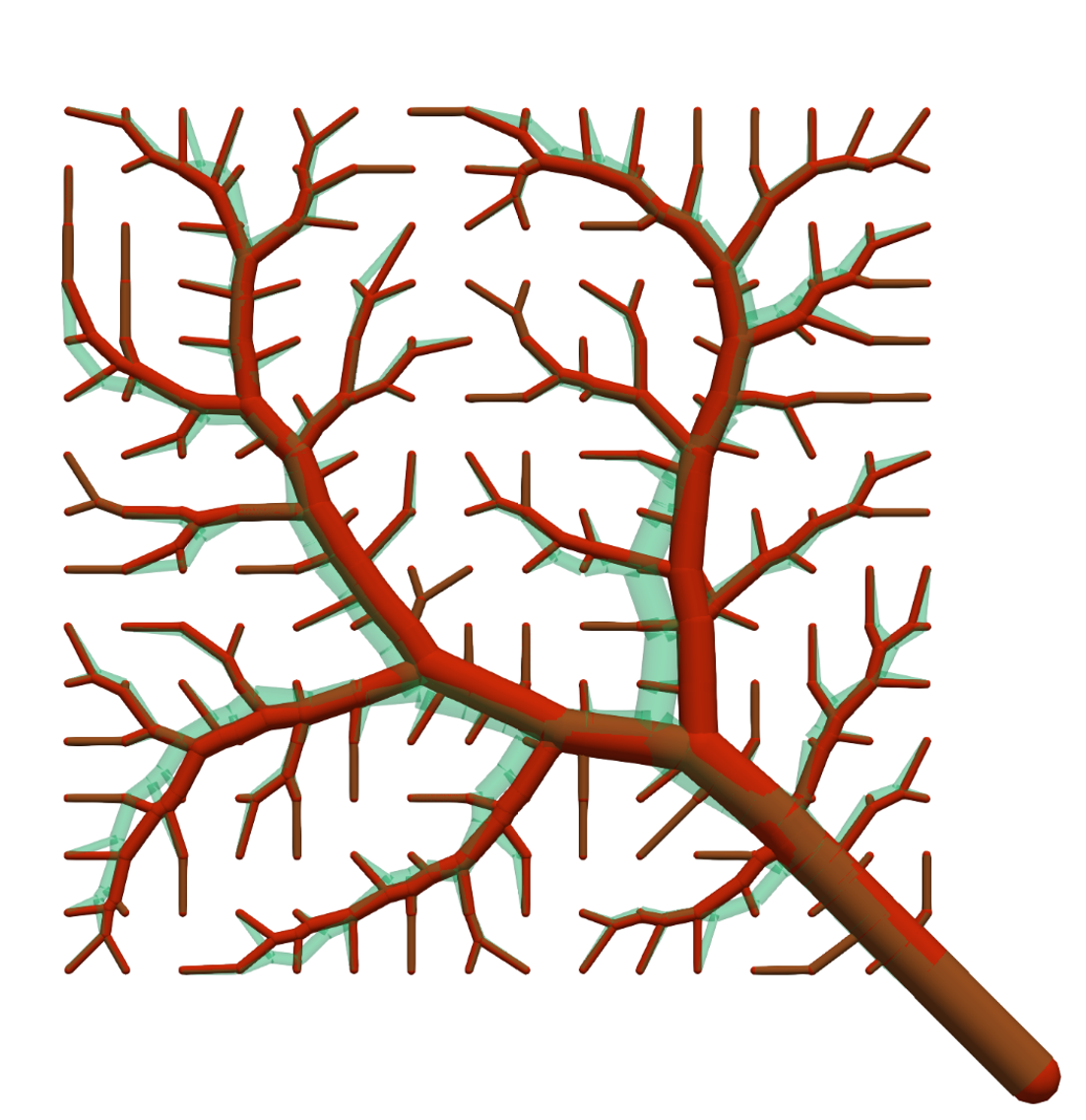}
  \caption{Comparison between local geometry optimization (green tree)
    and global geometry optimization (red tree) for $200$ terminal nodes.
    Optimizing the global geometry reduces the total cost by about $3$\%.}
  \label{fig:global_geometry_comparison}
\end{figure}

As described in \cite{jessen2022rigorous, jessen2023branching} in more detail,
we include all variables to be optimized
in the vector $y$.
For our baseline problem, this only includes
the nodal positions $x = (x_v)_{v \in \Nodes}$ and
the vessel lengths $\ell = (\ell_a)_{a \in \Arcs}$,
giving the variable vector $y = (x, \ell) \in \R^{3 \cardV + \cardA}$.
We add a physical lower bound $\ell^-$ and,
for numerical efficiency, an upper bound $\ell^+$.
Because we do not generate the microcirculation,
the lower bound $\ell^-$ is set to \SI{0.01}{\mm}.
The upper bound $\ell^+$ must be set based on the organ
and should allow the largest possible vessels to form.
Then, the best geometry is found in the rectangle defined as
\begin{align}
  Y = \R^{3 \cardV} \x [\ell^-, \ell^+]^{\cardA},
\end{align}
and the NLP reads:
\begin{align}
  \label{nlp-baseline}
  \min_{y \in Y} \quad
  & \sum_{a \in \Arcs} \ell_a w_a(r_a) \\
  \stq
  \label{eq:nlp_baseline-fix-x}
  &0 = x_u - \_x_u, & u &\in \set{0} \cup \Leaves, \\
  \label{eq:nlp_baseline-length}
  &0 = \ell_{uv}^2 - \norm{x_u - x_v}^2, & uv &\in \Arcs.
\end{align}
Here \eqref{eq:nlp_baseline-fix-x} fixes the positions
of the root node and the terminal nodes,
and \eqref{eq:nlp_baseline-length} ensures consistency
between nodal positions and segment lengths.
The effect of global geometry optimization
by means of our NLP is illustrated
in \cref{fig:global_geometry_comparison}
where we synthesized the tree inside a shallow rectangular box
with dimensions $\SI{100}{mm} \x \SI{100}{mm} \x \SI{20}{mm}$
and with the root node located outside the bottom right corner.
All figures show the projection to the $xy$-plane.
The total energy expenditure of the original tree
(based on \eqref{eq:murray-cost}) is reduced by around $3\%$
with our NLP approach.
Furthermore, the global optimization straightens
the ``jagged'' paths between segments,
which arise as a consequence of repeated local optimizations.

\subsubsection{Extension to nonconvex domains}
\begin{figure*}[ht]
  \centering
  \includegraphics[trim=0 0 10 0, clip=true,width=1.0\textwidth]{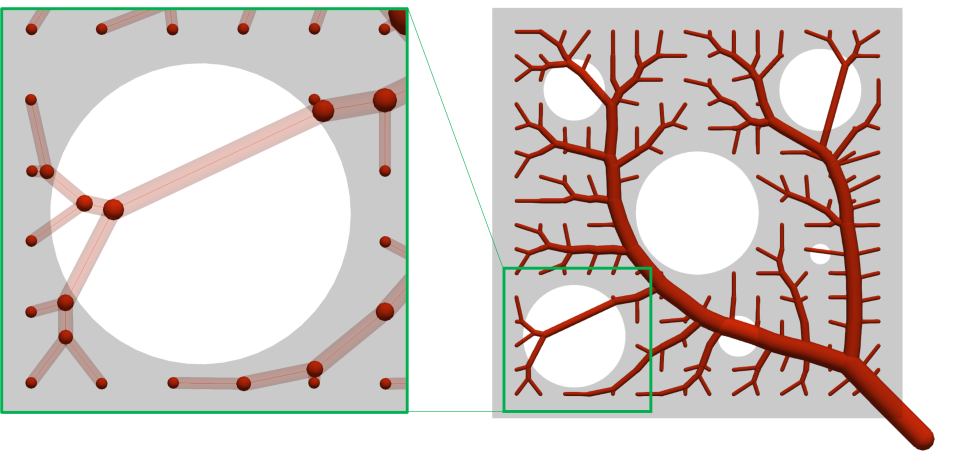}
  \caption{Vascular tree inside a nonconvex perfusion volume
    (grey, rectangular shape with six ``holes'').
    Left side: Even with terminal nodes sampled inside $\Omega$,
    branching nodes (red spheres) and edges (shaded connections)
    can be synthesized outside $\Omega$.}
  \label{fig:non_convex_volume}
\end{figure*}

The global geometry optimization works for any convex perfusion volume
and more generally when the shortest weighted
connections between two tree nodes are always inside $\Omega$.
However, nonconvex features such as holes
may require further effort,
as shown in \cref{fig:non_convex_volume}
for a rectangle-shaped perfusion volume $\Omega$ (in grey).
Even when terminal nodes are generated inside
the nonconvex perfusion volume,
both branching nodes and edges can lie partially or entirely
outside the domain $\Omega$
(highlighted on the left of \cref{fig:non_convex_volume}).

We assume that $\Omega$ is given by its boundary $\surface$,
a closed surface consisting of finitely many triangles.
Such a \emph{triangulation} is a standard discretization
of the physical volume's boundary.
It enables cheap computations of many geometric properties,
such as testing whether a given point belongs to $\Omega$
or determining its distance to $\surface$.
All the data sets that we consider in this work represent $\surface$
(and thus $\Omega$) by a sufficiently fine triangulation.
With this assumption for $\Omega$, we can generate all terminal nodes
uniformly inside $\Omega$ at the start of the algorithm.
We sample all points inside the bounding box of $\Omega$ on some regular grid
(with the addition of small random pertubations) and, subsequently,
delete each point outside $\Omega$.
By choosing the density of the grid based on the volume ratio
between bounding box and $\Omega$, the final sample size $N\term$
only deviates slightly from the desired number of terminals.

\begin{figure*}[ht]
  \centering
  \subfloat[$\veps$ = \SI{2.0}{\mm}]
  {\includegraphics[trim=0 0 0 0,clip=true,width=0.3\textwidth]
  {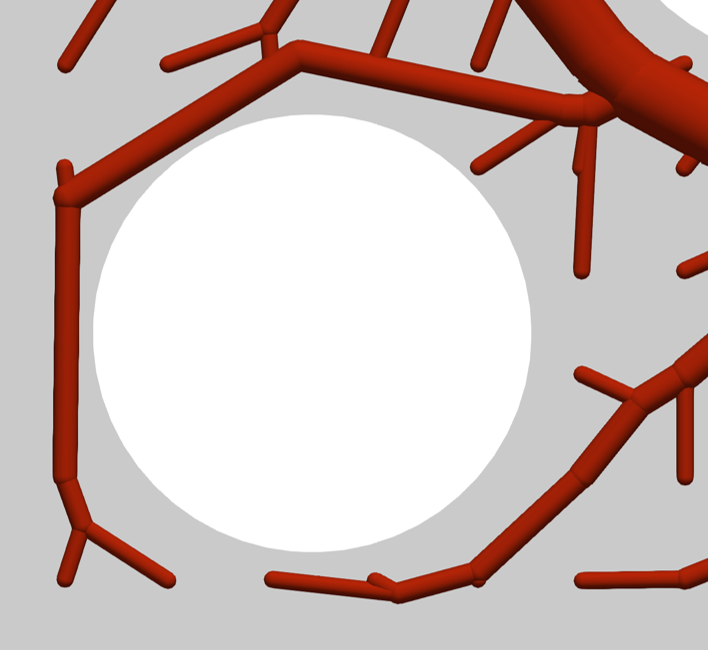}%
    }\label{fig:non_convex_tol_05}\hfill
  \subfloat[$\veps$ = \SI{1.0}{\mm}]
  {\includegraphics[trim=0 0 0 0,clip=true,width=0.3\textwidth]
  {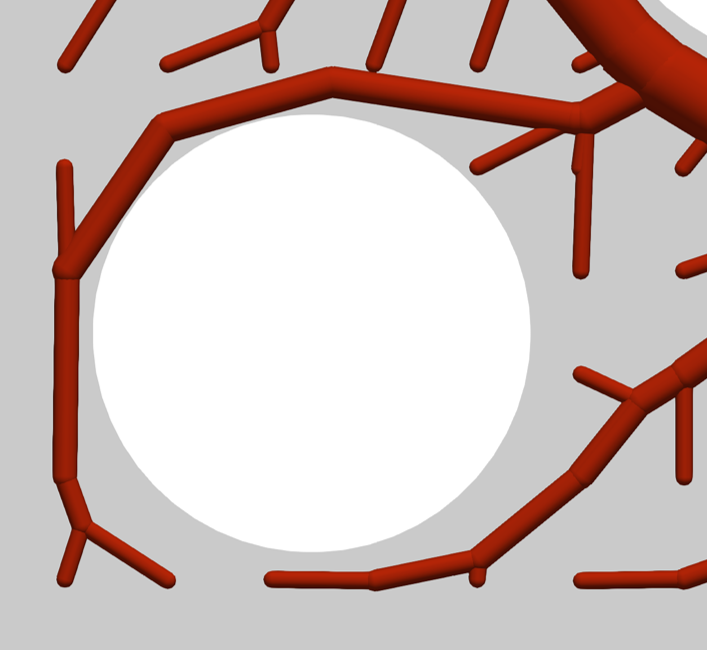}%
    }\label{fig:non_convex_tol_1}\hfill
  \subfloat[$\veps$ = \SI{0.5}{\mm}]
  {\includegraphics[trim=0 0 0 0,clip=true,width=0.3\textwidth]
  {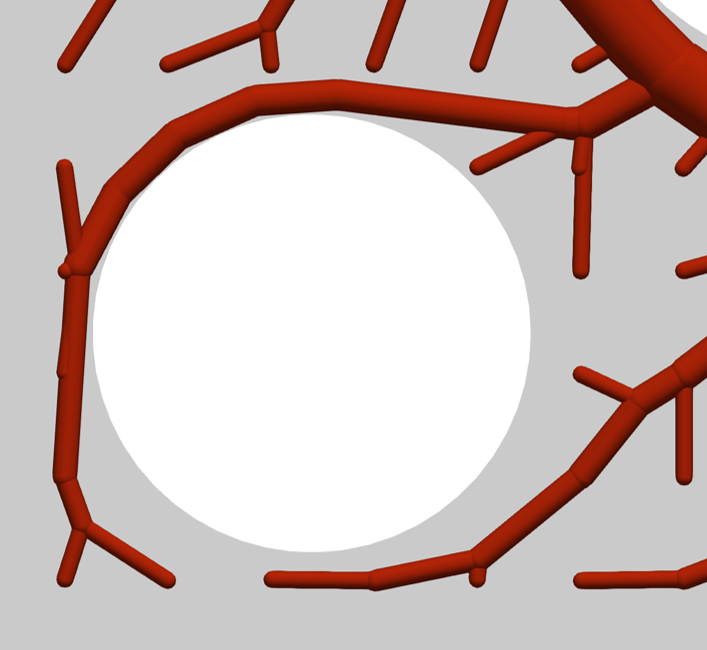}%
    }\label{fig:non_convex_tol_2}
  \caption{Influence of the allowed minimum distance $\veps$
    between balls and nodes of the tree.
    Smaller values of $\veps$ lead to more optimal geometry (tighter curves)
    at the expense of additional extension nodes.}
  \label{fig:non_convex_tolerance}
\end{figure*}

Now, given a nonconvex domain $\Omega$,
we first solve the NLP \eqref{nlp-baseline}--\eqref{eq:nlp_baseline-length}
to synthesize an initial vascular tree
that will typically not lie entirely in $\Omega$.
Instead of introducing local changes
to individually move nodes and segments inside $\Omega$,
our goal is then to achieve this by repeating our global geometry optimization
with additional constraints.
To this end, we compute the set $\extNodes$ of tree nodes outside $\Omega$
and the set $\extArcs$ of segments that pass through the exterior of $\Omega$
even though both its proximal and distal nodes are already inside.
For every node $v \in \extNodes$,
we set $c_v \define x_v$ (the node's infeasible location)
and $d_v \define \dist(c_v, \surface)$ (its distance to the boundary).
Then we extend the original NLP with a constraint
that exludes the open ball
$B(c_v, d_v + \veps_v) \define \defset{x}{\norm{x - c_v} < d_v + \veps_v}$
from the feasible region for $x_v$,
where the parameter $\veps_v$ is chosen to be larger
than the largest segment radius at node~$v$.
For every segment $a = uw \in \extArcs$,
we determine the point $c_a$ on the line segment $[x_u, x_w]$
that has the largest distance $d_a$ to $\surface$.
Then we add an \emph{extension node} $v \equiv v_a$
(originally located at $c_a$)
to split the segment $a = uw$
into two parts $uv$ and $vw$ that may form a kink,
and we extend the original NLP with a constraint
that excludes the open ball $B(c_a, d_a + \veps_a)$
from the feasible region for $x_v$
and with constraints that prevent a shortening of the two new segments.
The parameter $\veps_a$ is chosen to be larger than the segment radius $r_a$.
We denote by $\setE \define \defset{v_a}{a \in \extArcs}$
the set of extension nodes
and by $\setB \define \extNodes \cup \setE$
the set of nodes with ball constraints and by $\Arcs_\setE \define \bigcup_{uw \in \extArcs} \set{uv_a, v_aw}$
the set of split arc pairs.
For $v \equiv v_a \in \setE$ with $a = uw$
we write $c_v, d_v, \veps_v$ instead of $c_a, d_a, \veps_a$,
and we set $\_\ell_{uv} \define \norm{x_u - c_v}$
and $\_\ell_{vw} \define \norm{c_v - x_w}$.
Our variable vector $y$ now also includes
the location $x_v$ of each extension node,
its distance $\ell_v$ to the associated ball center $c_v$,
and the lengths $\ell_{uv}, \ell_{vw}$ of split arc pairs.
The best geometry (inside the nonconvex volume) is then found in
\begin{align}
  Y = \R^{3 \cardN + 3 \card\setE + \card\setB}
  \x [\ell^-, \ell^+]^{\cardA + \card\extArcs + \card\setB},
\end{align}
and the extended NLP reads:
\begin{align}
  \label{nlp-nonconvex}
  \min_{y \in Y} \quad
  &\sum_{a \in \Arcs} \ell_a w_a(r_a) \\
  \stq
  &0 = x_u - \_x_u, & u &\in \set{0} \cup \Leaves, \\
  &0 = \ell_{uv}^2 - \norm{x_u - x_v}^2,
  & uv &\in \Arcs \setminus \extArcs \cup \Arcs_\setE, \\
  \label{eq:nlp-nonconvex-length}
  &0 = \ell_{v}^2 - \norm{x_v - c_{v}}^2, & v &\in \setB, \\
  \label{eq:nlp-nonconvex-radius}
  &\ell_{v} \ge d_v + \veps_v, & v &\in \setB, \\
  \label{eq:nlp-excursion-distal-length}
  &\ell_{uv} \ge \_\ell_{uv}, \ \ell_{vw} \ge \_\ell_{vw},
  & v &\in \setE,\ uw \in \extArcs.
\end{align}
Clearly, \eqref{eq:nlp-nonconvex-length} defines the length $\ell_v$
and \eqref{eq:nlp-nonconvex-radius} forces each node in $\setB$
outside its corresponding ball (by a tolerance distance $\veps_v$).
Smaller values of $\veps_v$ lead to more ``natural''
(near-optimal) vessel curvature
while increasing the number of extension nodes
that need to be added,
as highlighted in \cref{fig:non_convex_tolerance}
for a common value of $\veps$ at all nodes.
The two last constraints
\eqref{eq:nlp-excursion-distal-length}
prevent extension nodes from moving closer
to either their proximal or distal node and, instead,
to form a kink while moving
outside their corresponding ball.
\\
\\
\textbf{Remark 6.1:}
The radius of a segment $uw$ only depends
on the number of leave nodes downstream of node~$w$.
Thus, adding an extension node $v$ on segment $uw$
does not change the radii for the new segments $uv$ and $vw$,
which are both equal to $r_{uw}$.
\\

\begin{figure*}[ht]
  \centering
  \subfloat[Detection of two external nodes (green and purple)
      and their corresponding balls]
  {\includegraphics[trim=0 0 0 0,clip=true,width=0.3\textwidth]
  {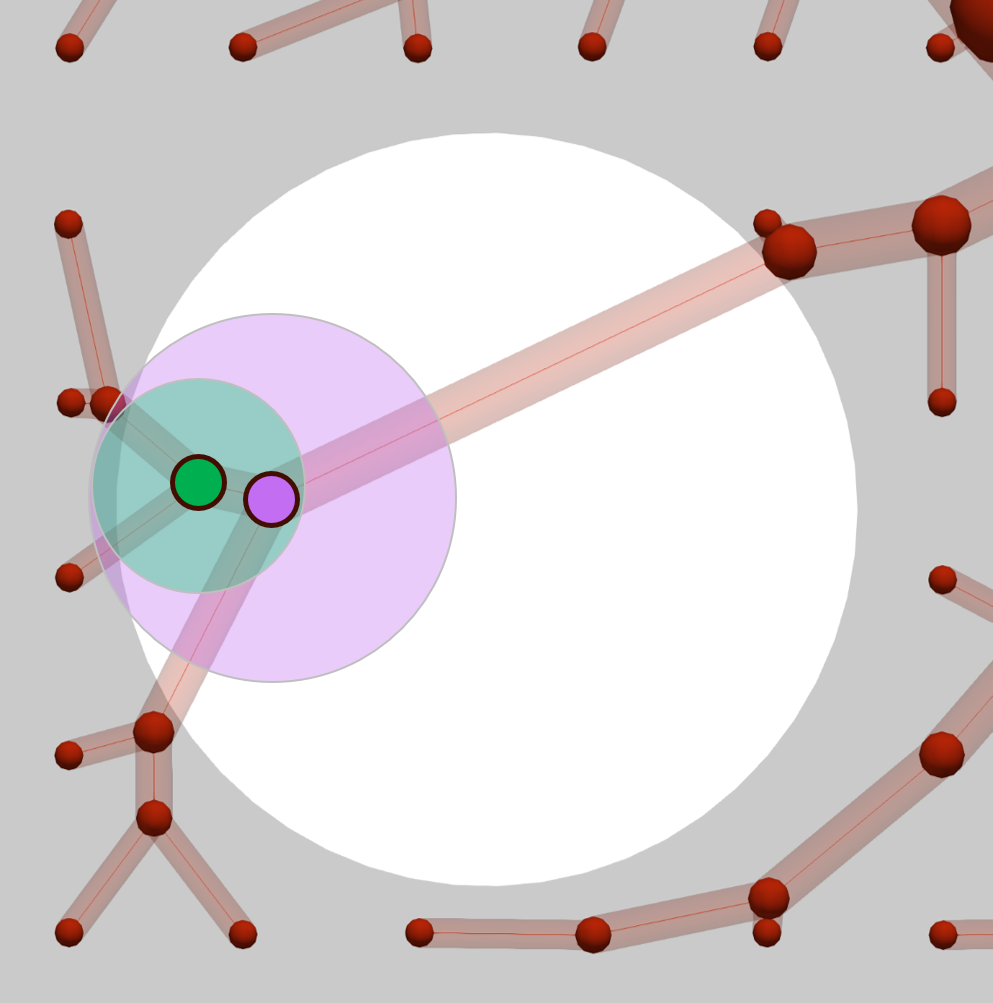}%
    }\label{fig:non_convex_iter_1}\hfill
  \subfloat[Reoptimization using extended NLP (yielding $\extArcs \ne \0$)]
  {\includegraphics[trim=0 0 0 0,clip=true,width=0.3\textwidth]
  {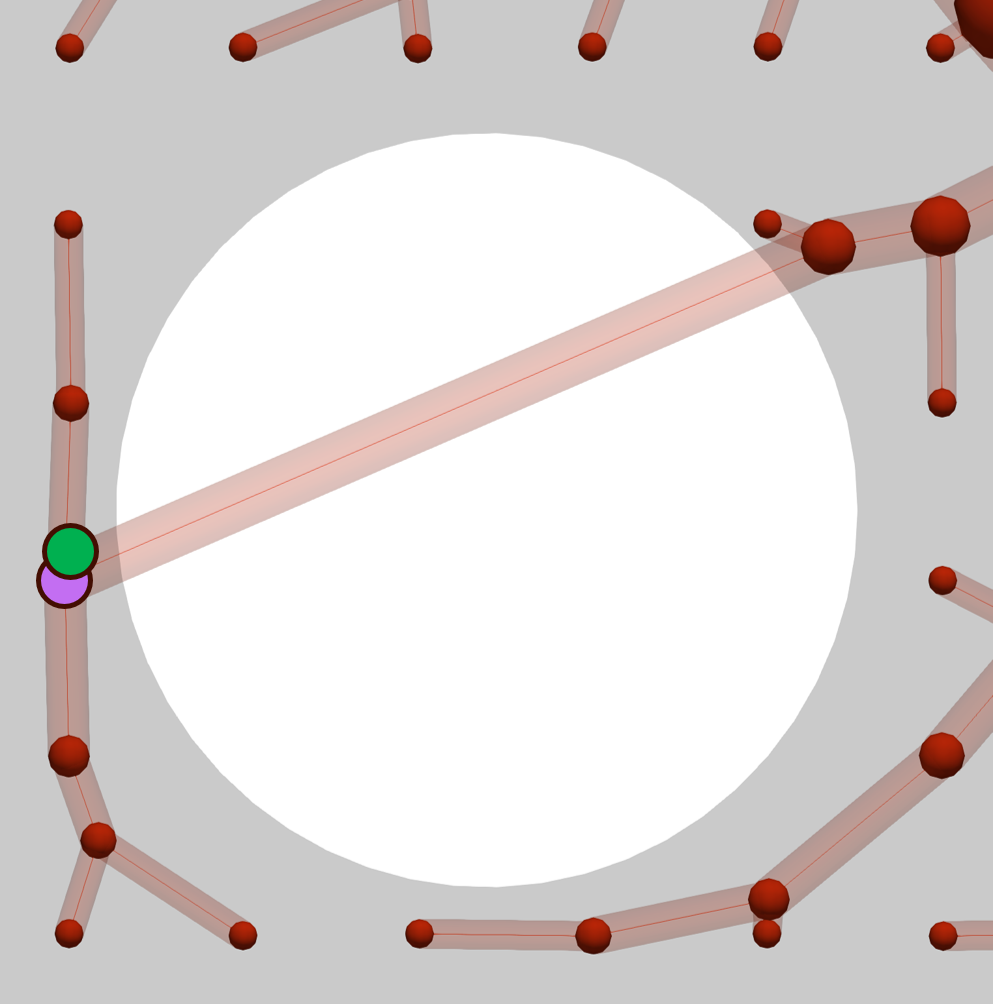}%
    }\label{fig:non_convex_iter_2}\hfill
  \subfloat[Adding one extension node (orange)
      and its corresponding ball]
  {\includegraphics[trim=0 0 0 0,clip=true,width=0.3\textwidth]
  {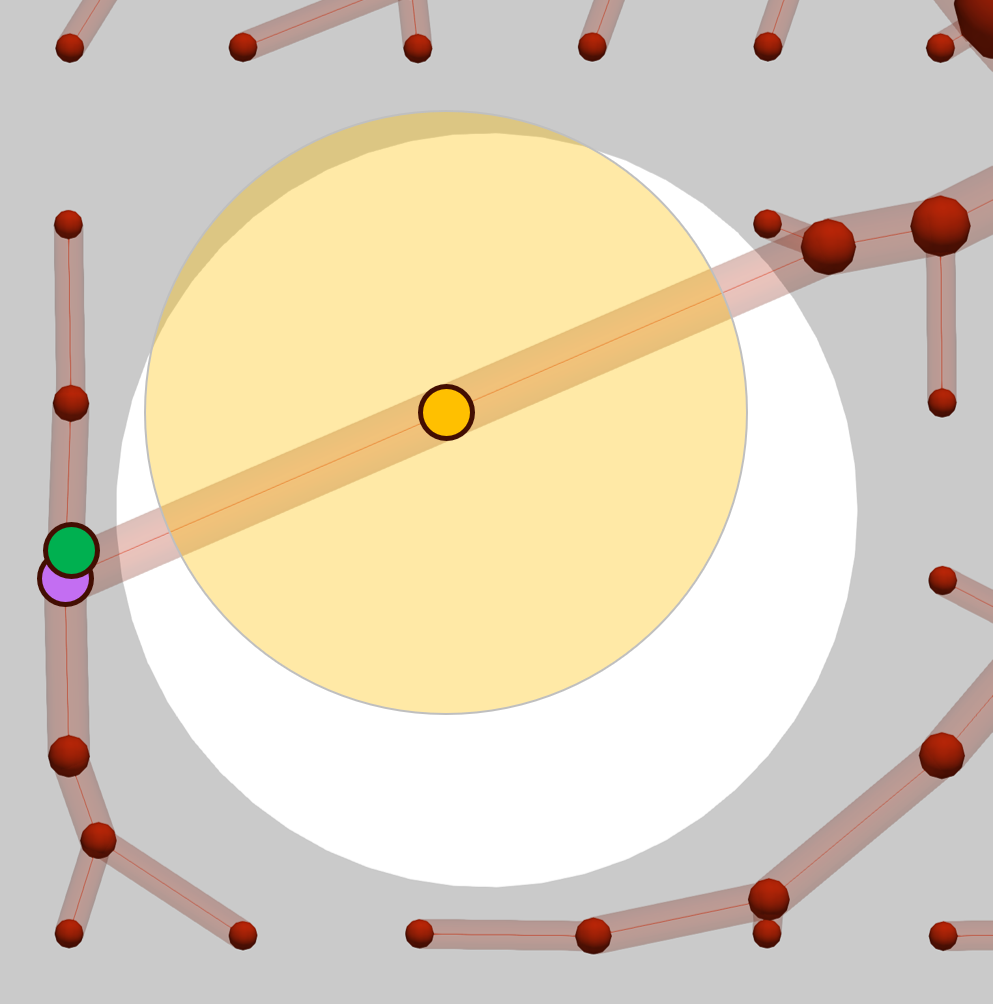}%
    }\label{fig:non_convex_iter_3}\\
    \subfloat[Reoptimization using extended NLP (yielding $\extArcs \ne \0$)]
  {\includegraphics[trim=0 0 0 0,clip=true,width=0.3\textwidth]
  {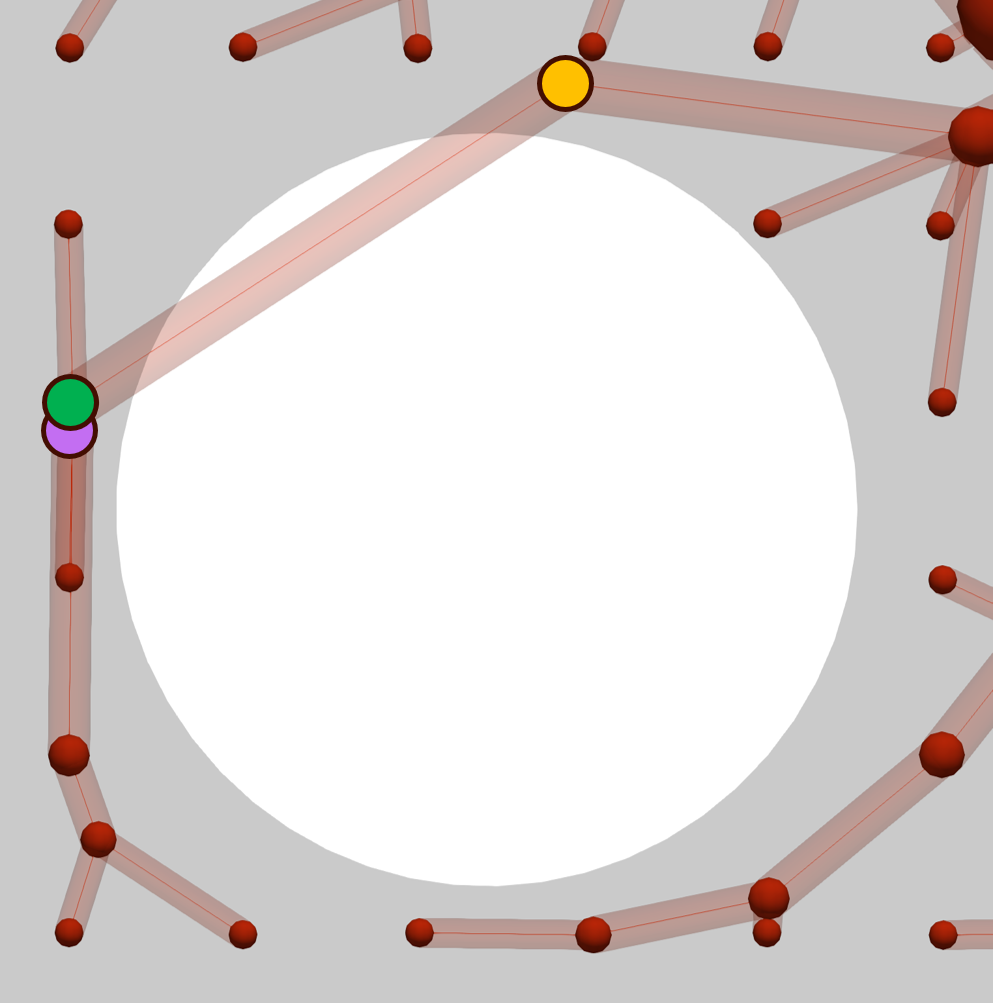}%
    }\label{fig:non_convex_iter_4}\hfill
  \subfloat[Adding one extension node (blue)
      and its corresponding ball]
  {\includegraphics[trim=0 0 0 0,clip=true,width=0.3\textwidth]
  {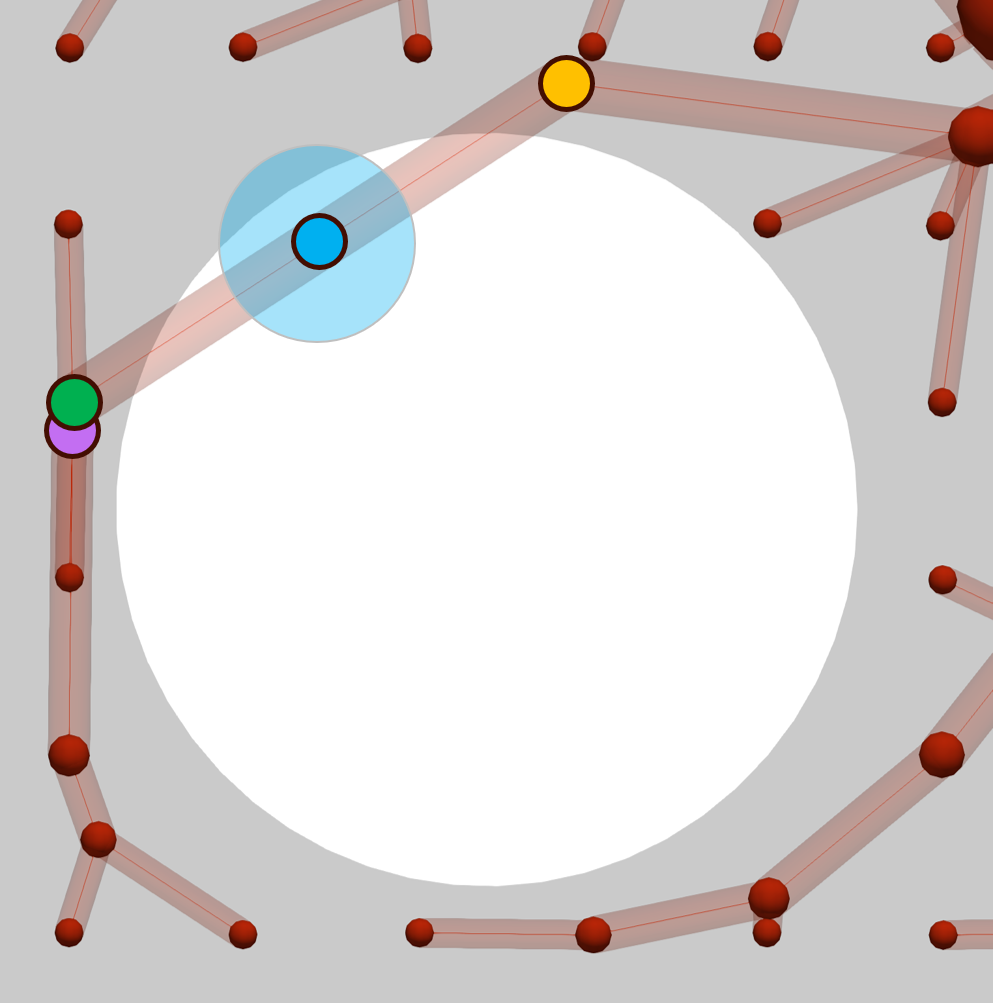}%
    }\label{fig:non_convex_iter_5}\hfill
  \subfloat[Reoptimization using extended NLP
  (yielding $\extNodes = \0$ and $\extArcs = \0$)]
  {\includegraphics[trim=0 0 0 0,clip=true,width=0.3\textwidth]
  {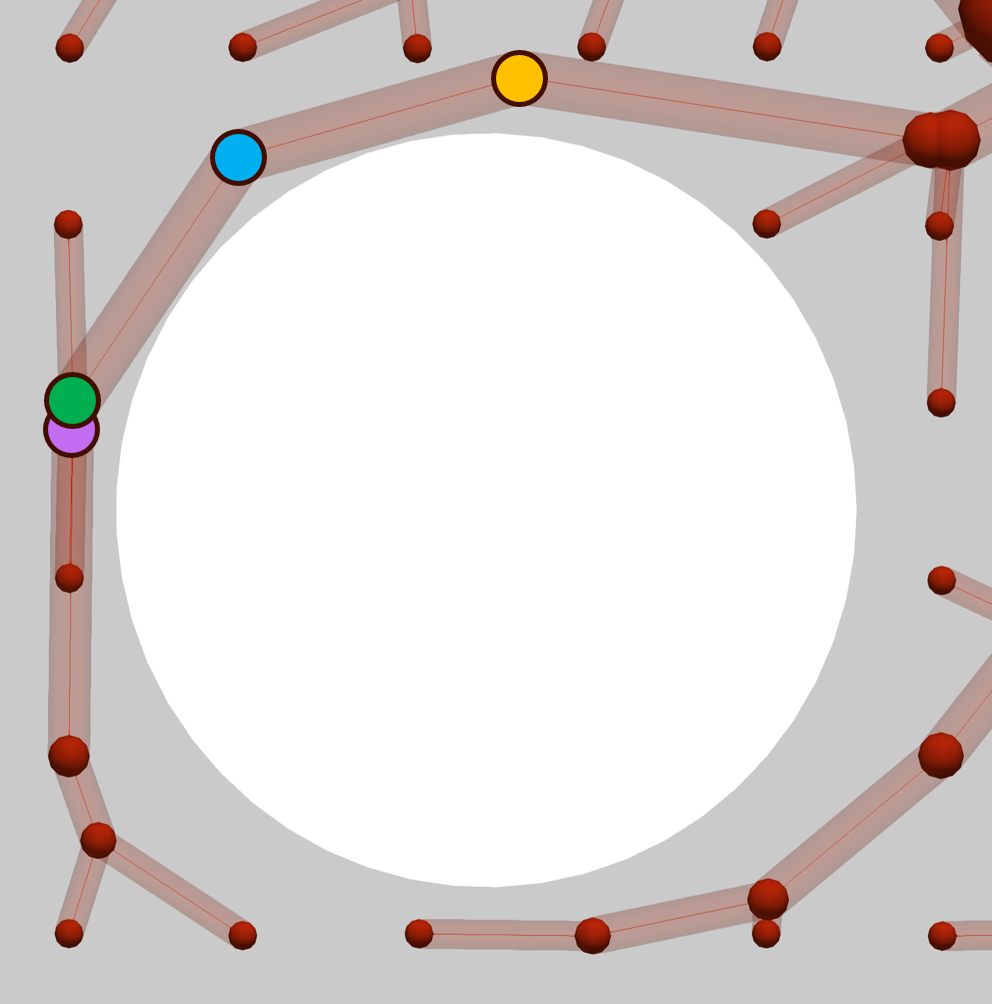}%
    }\label{fig:non_convex_iter_6}
  \caption{Steps to constrain a vascular tree
    inside a nonconvex perfusion volume $\Omega$.
    To resolve this section of $\Omega$,
    a total of two new extension nodes are added.}
  \label{fig:non_convex_iterations}
\end{figure*}

The (general) detection of nodes outside the perfusion volume
and the computation of the initial extension locations
can not be formulated as NLP constraints.
Therefore these operations must be
executed outside (before) solving the NLP.
The result is an iterative approach
wherein we repeatedly detect and add new extensions
to optimize the global geometry
constrained to the nonconvex perfusion volume
until all segments lie entirely inside $\Omega$.
This process is visualized in \cref{fig:non_convex_iterations} and
the complete algorithm is summarized in \cref{alg:nonconvex}.

\begin{algorithm}[ht]
  \caption{Nonconvex algorithm}
  \label{alg:nonconvex}
  \begin{algorithmic}[1]
    \State Compute an initial tree using the NLP
    \eqref{nlp-baseline}--\eqref{eq:nlp_baseline-length}
    \State Compute the sets $\extNodes$ and $\extArcs$
    \While{$\extNodes \ne \0$ or $\extArcs \ne \0$}
    \State Add extension nodes to the segments in $\extArcs$
    \State Identify the set of nodes with ball constraints $\setB$
    \State Reoptimize the tree geometry using the NLP
    \eqref{nlp-nonconvex}-\eqref{eq:nlp-excursion-distal-length}
    \State Recompute the sets $\extNodes$ and $\extArcs$
    \EndWhile
  \end{algorithmic}
\end{algorithm}

\subsubsection{Extension to multiple non-intersecting trees}
\label{sec:intersection_fix}
\begin{figure}[ht]
  \centering
  \includegraphics[width=0.5\textwidth]{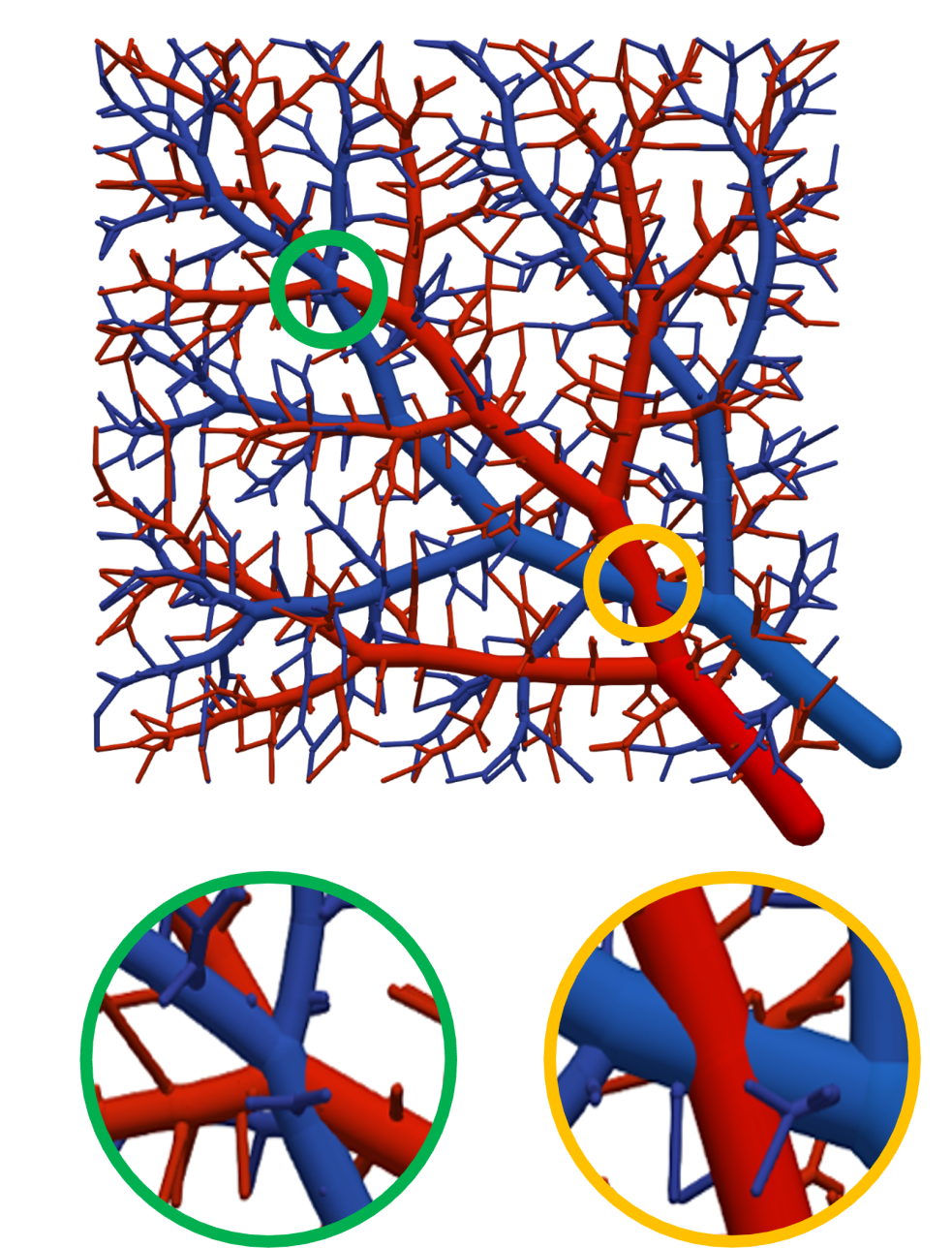}
  \caption{Two vascular trees generated in the same perfusion volume $\Omega$.
    Without additional constraints, these trees can penetrate each other.
    Two intersections are highlighted in green and orange.
    The two trees intersect a total of $31$ times (total tree energy 4334).}
  \label{fig:intersecting_trees}
\end{figure}

The perfusion of an organ
involves at least two vascular trees
(with common terminal nodes);
typically one inlet and one outlet tree.
However, without additional constraints,
two generated trees inside the same perfusion domain
face a high risk of intersecting each other,
as highlighted in \cref{fig:intersecting_trees}.
Similar to the treatment of nonconvexity,
our proposed approach for removing any intersections consists in
synthesizing multiple vascular trees independently in a first step
and repeating our NLP-based global geometry optimization
with appropriate additional constraints.
Every reoptimization step in this process
combines the global geometry models of all trees
along with all added non-intersection constraints
in a single extended NLP.

\begin{figure}[ht]
  \centering
  \includegraphics[width=0.55\textwidth]{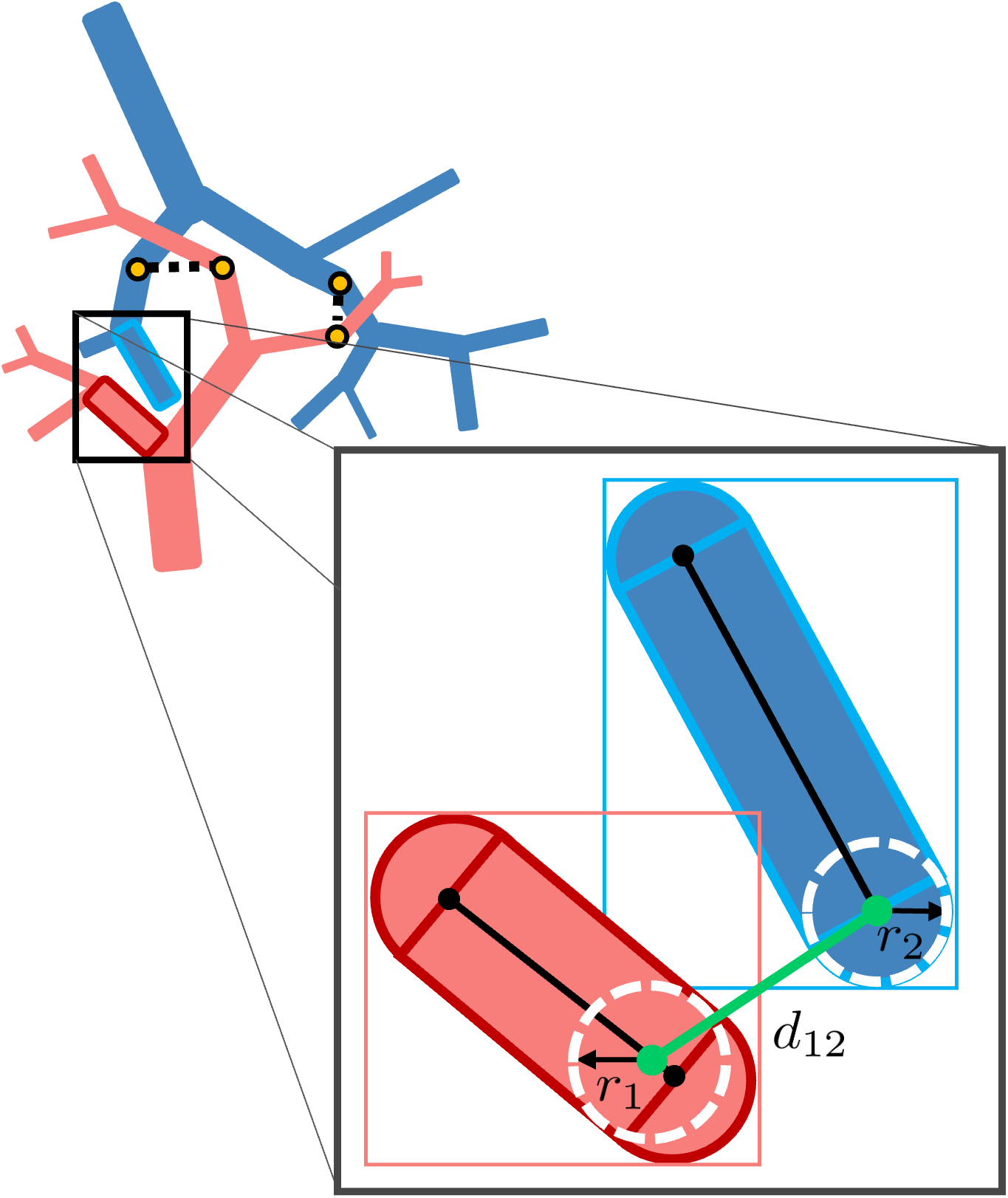}
  \caption{Steps to identify and remove intersecting vessels.
    Bounding boxes and capsules are used to find intersections.
    New extension nodes (orange) are introduced to remove these intersections.}
  \label{fig:intersection_test}
\end{figure}

To identify intersections efficiently,
we treat each segment as a capsule,
i.e., as a cylinder with a half-ball attached to each end;
see \cref{fig:intersection_test}.
The intersection test between two segments then consists of two parts.
The first is finding the pair of nearest points
on the two cylinder axes (line segments).
The second part is to test whether the distance between these
two nearest points is smaller than the combined radii of both capsules.
We further accelerate the identification of intersections
by utilizing appropriate bounding boxes.
Based on this identification, \cite{guy20193d} proposed a local algorithm
to remove intersections by introducing extension nodes
at the locations of intersections
and moving them apart based on their intersection depth.
This algorithm can remove any intersection,
but it introduces a large number of extension nodes,
even for larger vessels.
The result is a significant increase in metabolic cost
and an unnaturally high curvature of modified vessels.

By allowing all nodes to move,
both the number of extension nodes
and the increase in cost should be significantly reduced
with our reoptimization approach.
Since the first part of the intersection test can not be
formulated as NLP constraints,
we can only include the second part.
Therefore the identification of the closest points
(and addition as extension nodes)
must be performed before solving the NLP,
which results in an iterative approach
similar to \cref{alg:nonconvex}.

For the reoptimization, we consider a \emph{forest} consisting of
any number $N_\Tree$ of trees $\Tree_i = (\Nodes_i, \Arcs_i)$,
represented by its index set
\begin{equation}
  \setF = \set{1, \dots,\ N_\Tree}.
\end{equation}
We check for intersections inside the forest pairwise with
\begin{equation}
  \setP = \defset{(i, j)}{i, j \in \setF,\ i < j}
\end{equation}
and introduce
for each tree pair $(i, j) \in \setP$
the set of all pairs $(a_i, a_j)$ of intersecting segments
along with their associated nearest points $(c_i, c_j)$,
\begin{equation}
  \Arcs_{ij} = \defset{(a_i, a_j, c_i, c_j)}
  {(a_i, a_j) \in \Arcs_i \x \Arcs_j \text{ intersect}}.
\end{equation}
Every quadruple in $\Arcs_{ij}$ with $a_i = u_iw_i$ and $a_j = u_jw_j$
therefore has $c_i \in [u_i, w_i]$ and $c_j \in [u_j, w_j]$ such that
\begin{equation}
  \dist([u_i, w_i], [u_j, w_j]) = \norm{c_i - c_j} < r_{a_i} + r_{a_j}.
\end{equation}
We define parameters $\_\ell_{u_ic_i} \define \norm{x_{u_i} - c_i}$
and $\_\ell_{c_iw_i} \define \norm{c_i - x_{w_i}}$,
and similarly for $c_j$.
Our variable vector~$y$ now includes
the variables of all individual trees as well as the
locations $x_{v_i}, x_{v_j}$ of the extension nodes
(initially set to $c_i, c_j$)
with associated lengths $\ell_{v_i}, \ell_{v_j}$,
and the lengths $\ell_{u_iv_i}, \ell_{v_iw_i}$ of split arc pairs.
Using the union $\Arcs_\setP \define \bigcup_{(i,j) \in \setP} \Arcs_{ij}$,
the extended NLP then becomes:
\begin{align}
  \label{nlp-no-intersection}
  \min_{y \in Y} \quad
  &\sum_{i \in \setF} \sum_{a_i \in \Arcs_i} \ell_{a_i} w_{a_i}(r_{a_i}) \\
  \stq
  &0 = x_{u_i} - \_x_{u_i},
  &&i \in \setF,\ u_i \in \set{0}_i \cup \Leaves, \\
  &0 = \ell_{u_iv_i}^2 - \norm{x_{u_i} - x_{v_i}}^2,
  &&i \in \setF,\ u_iv_i \in \Arcs_i, \\
  \label{nlp-virtual-length}
  &0 = \ell_{v_iv_j}^2 - \norm{x_{v_i} - x_{v_j}}^2,
  &&(a_i,a_j,v_i,v_j) \in \Arcs_{\setP}, \\
  \label{nlp-virtual-length-constraint}
  &\ell_{v_iv_j} \ge r_{a_i} + r_{a_j} + \veps,
  &&(a_i,a_j,v_i,v_j) \in \Arcs_{\setP}, \\
  \label{eq:nlp-excursion-proximal-length2}
  &\ell_{u_iv_i} \ge \_\ell_{u_ic_i},\ \ell_{u_jv_j} \ge \_\ell_{u_jc_j},
  &&(a_i,a_j,v_i,v_j) \in \Arcs_{\setP}, \\
  \label{eq:nlp-excursion-distal-length2}
  &\ell_{v_iw_i} \ge \_\ell_{c_iw_i},\ \ell_{v_jw_j} \ge \_\ell_{c_jw_j},
  &&(a_i,a_j,v_i,v_j) \in \Arcs_{\setP}.
\end{align}

\begin{figure*}[ht]
  \centering
  \subfloat[Initial intersecting trees]
  {\includegraphics[trim=0 150 0 50,clip=true,width=0.43\textwidth]
  {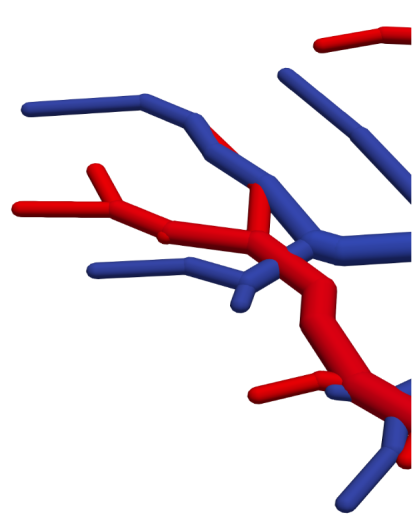}%
    }\label{fig:no_intersect_iter_1}\hfill
  \subfloat[Add extension nodes (iter.~1)]
  {\includegraphics[trim=0 150 0 50,clip=true,width=0.43\textwidth]
  {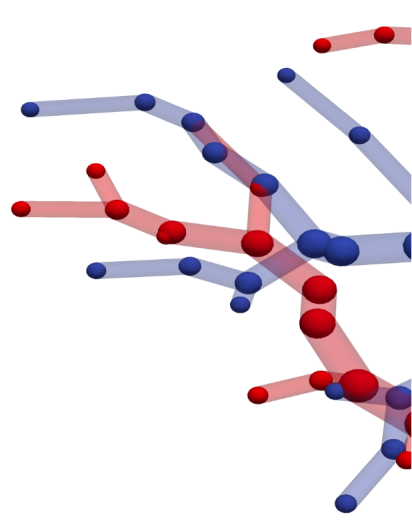}%
    }\label{fig:no_intersect_iter_2}\\
  \subfloat[NLP solution (iter.~1)]
  {\includegraphics[trim=0 150 0 50,clip=true,width=0.43\textwidth]
  {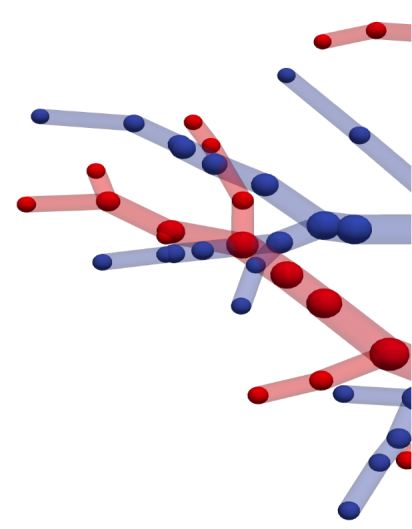}%
    }\label{fig:no_intersect_iter_3}\hfill
  \subfloat[Add extension nodes (iter.~2)]
  {\includegraphics[trim=0 150 0 50,clip=true,width=0.43\textwidth]
  {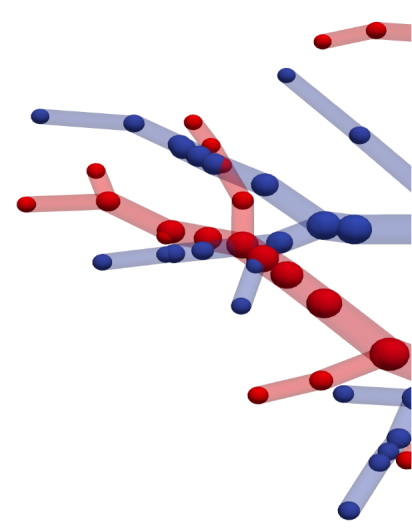}%
    }\label{fig:no_intersect_iter_4}\\
    \subfloat[NLP solution (iter.~2)]
  {\includegraphics[trim=0 150 0 50,clip=true,width=0.43\textwidth]
  {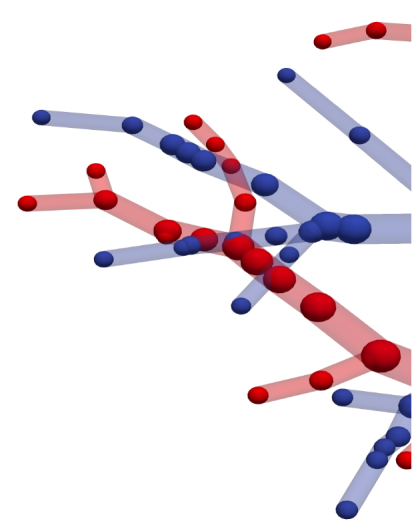}%
    }\label{fig:no_intersect_iter_5}\hfill
  \subfloat[Add extension nodes (iter.~3)]
  {\includegraphics[trim=0 150 0 50,clip=true,width=0.43\textwidth]
  {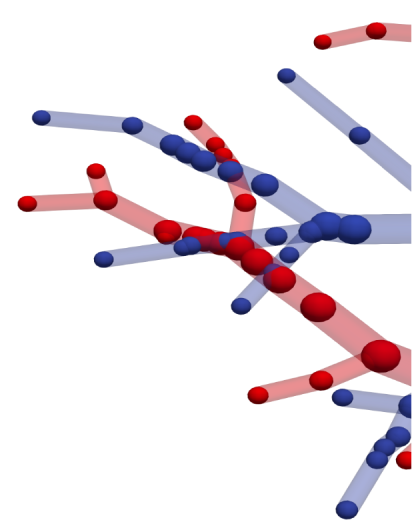}%
    }\label{fig:no_intersect_iter_6}\\
  \subfloat[NLP solution (iter.~3)]
  {\includegraphics[trim=0 150 0 50,clip=true,width=0.43\textwidth]
  {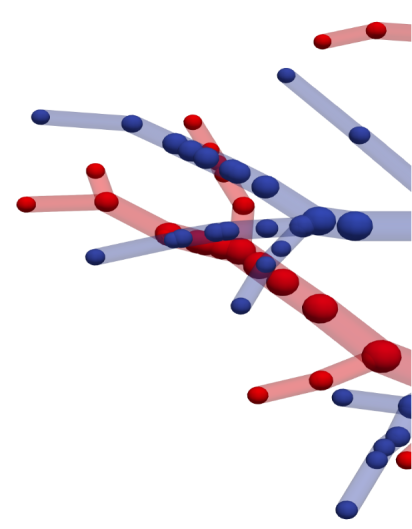}%
    }\label{fig:no_intersect_iter_7}\hfill
  \subfloat[Final non-intersecting trees]
  {\includegraphics[trim=0 150 0 50,clip=true,width=0.43\textwidth]
  {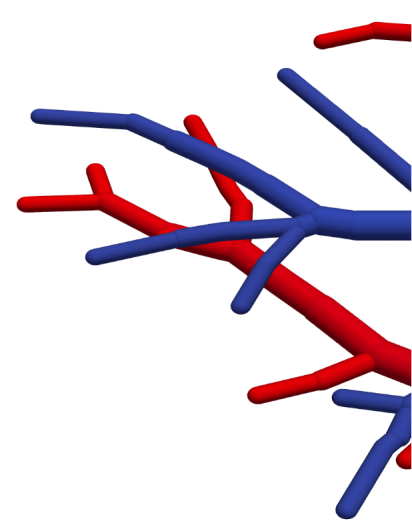}%
    }\label{fig:no_intersect_iter_8}
  \caption{Steps to (globally) remove intersections between two vascular trees.}
  \label{fig:no_intersection_iterations}
\end{figure*}

Herein \eqref{nlp-virtual-length} defines the distance between
a pair of extension nodes of two trees
and \eqref{nlp-virtual-length-constraint} ensures that the
associated segments do not overlap
at these nodes (with a tolerance~$\veps$).
As for the nonconvex volume,
\eqref{eq:nlp-excursion-proximal-length2} and
\eqref{eq:nlp-excursion-distal-length2}
force segments with extension nodes to develop kinks.
To completely remove all intersections,
we need to iterate between adding new extension nodes at intersections
and solving the NLP, as shown in \cref{alg:no-intersection}.
\Cref{fig:no_intersection_iterations} highlights
how this approach iteratively removes all intersections.
\begin{algorithm}[ht]
  \caption{No-intersection algorithm}
  \label{alg:no-intersection}
  \begin{algorithmic}[1]
    \State Compute $N_{\Tree}$ initial trees independently
    \State Detect and count the number of intersections $N$
    \While{$N > 0$}
    \State Add extension nodes to intersecting segments
    \State Identify the set $\Arcs_{ij}$ for each tree pair in the forest
    \State Reoptimize the tree geometry using the NLP
    \eqref{nlp-no-intersection}--\eqref{eq:nlp-excursion-distal-length2}
    \State Detect and count the number of intersections $N$
    \EndWhile
  \end{algorithmic}
\end{algorithm}

\subsection{Complete framework}
We combine our geometry optimization algorithms
with the SA algorithm, introduced in \cite{jessen2022rigorous},
into a unified generation framework
that is capable of generating vascular trees up to the micro-circulation.
The generation of the trees starts
by sampling the $N\term$ common terminal nodes
inside $\Omega$ based on the choice of distribution.
From this set of terminal nodes, $N\topo$ nodes are randomly sampled
and the topology of each tree is optimized
(with the same $N\topo$ nodes)
as described in \cite{jessen2022rigorous}.
The geometry of the resulting locally optimal topology
is then globally optimized and constrained to adhere to
(possibly) nonconvex perfusion volumes and multiple trees,
according to \cref{alg:nonconvex} and \cref{alg:no-intersection}.
Afterward, the number of current terminal nodes is doubled
(sampled from the initial $N\term$ terminal nodes).
To avoid unnecessary intersections,
we reject samples inside already existing vessels.
From here, the SA algorithm starts again.

During our tests, this process resulted in a good balance
between computational complexity and optimality of the final trees.
After all terminals are connected,
the trees can be reoptimized again
to adhere to more complicated physiological phenomena,
such as non-Newtonian blood rheology \cite{jessen2023branching}.
Lastly, any segments that reach the lower length bound $\ell^-$
are considered degenerate and are removed
and replaced with their branch segments,
possibly creating $n$-furcations with $n \ge 3$.
This complete generation framework is shown in \cref{alg:generation-framework}.

\clearpage
\begin{algorithm}[ht]
  \caption{Complete generation framework}\label{alg:generation-framework}
  \begin{algorithmic}[1]
    \State Sample $N\term$ terminal nodes inside $\Omega$
    \State Create $N_\Tree$ initial trees
    sharing $N = N\topo$ terminal nodes
    \While{$N < N\term$}
    \State Optimize trees using SA algorithm of \cite{jessen2022rigorous}
    \State Apply \cref{alg:nonconvex}
    and/or \cref{alg:no-intersection} (as needed)
    \State Set $M = \min(2N, N\term)$
    \For {$i = N+1, \ldots, M$}
    \If{node $x{\term}_{,i}$ outside existing vessels}
    \State Connect node $x{\term}_{,i}$ to nearest edge
    \EndIf
    \EndFor
    \State Set $N = M$
    \EndWhile
    \State Apply \cref{alg:nonconvex}
    and/or \cref{alg:no-intersection} (as needed)
    \State Replace degenerate segments with proximal node
  \end{algorithmic}
\end{algorithm}

\section{Results}\label{sec:results}
\subsection{Implementation details}
Our framework is implemented
in the programming language \emph{Julia} \cite{bezanson2017julia}.
The NLPs are solved using an interior point method
implemented in the program \emph{Ipopt} \cite{wachter2006implementation}
together with the linear solver \emph{Mumps}\cite{Amestoy_et_al:2001}.
All computations, including the reference benchmarks,
were done on a desktop computer with $64$ GB of random-access memory (RAM)
and an AMD Ryzen 7950x @5Ghz CPU with $32$ processing threads.

\subsection{Rectangle domain with holes}
\begin{figure}[ht]
  \centering
  \includegraphics[width=0.5\textwidth]{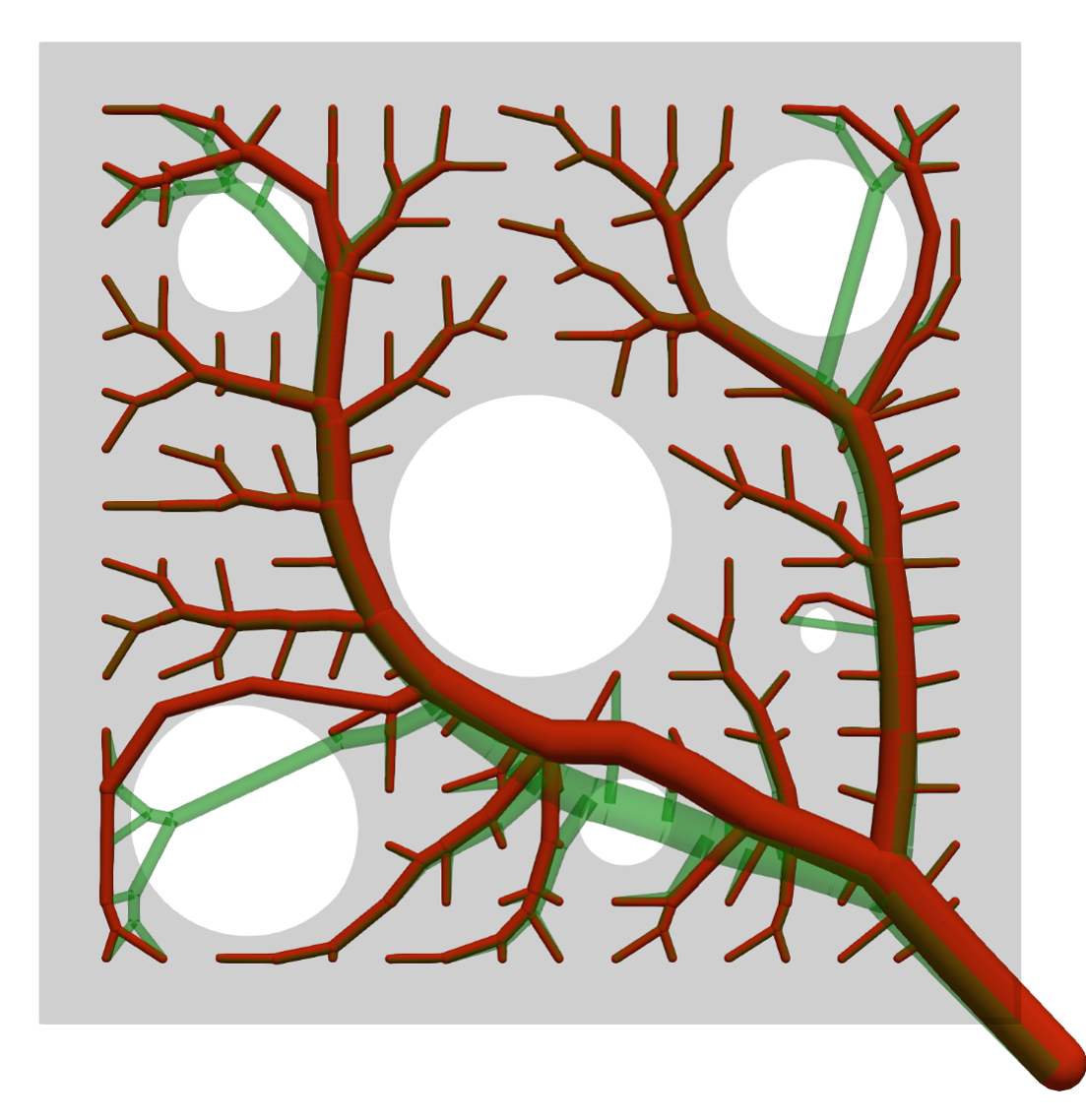}
  \caption{Vascular tree in red after optimization with additional
    constraints to keep the tree inside $\Omega$.
    The initial (unconstrained) tree is shown in green.}
  \label{fig:non_convex_tree}
\end{figure}

We start by restricting a single vascular tree to the nonconvex domain
introduced with \cref{fig:non_convex_volume}.
The excluded domain consists of six differently sized holes,
with diameters ranging from $\SI{2}{mm}$ to $\SI{25}{mm}$.
The resulting constrained tree is shown in \cref{fig:non_convex_tree}
in red against the initial unconstrained tree in green.
All vessels and nodes are moved inside the domain (grey).
The largest difference in geometry to the initial tree
is in the vicinity of the holes.
Because we consider the global geometry,
we also see differences in other areas of the tree.
Most notably, the vessels of the left trunk
behind the first major branch point moved slightly upwards
to alleviate the following kink near the bottom center hole.
The energy cost of the tree increased by $2.5\%$
with a total of $8$ extension nodes introduced.

\subsection{Two trees inside rectangle}
\begin{figure*}[ht]
  \centering
  \includegraphics[trim=25 0 10 0, clip=true,width=0.980\textwidth]{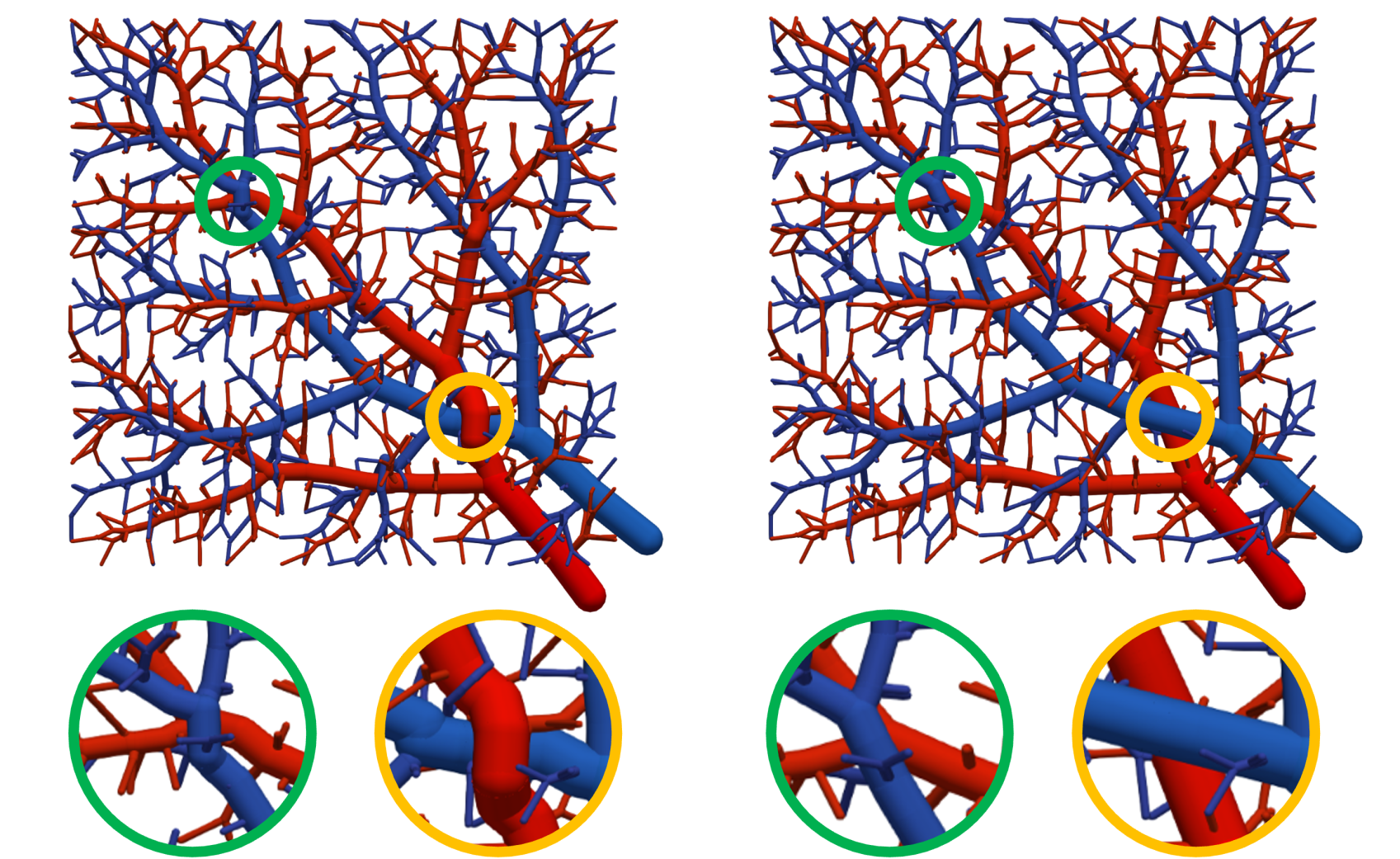}
  \caption{Comparison of the local intersection removal (left)
    and our new global algorithm (right). Both algorithms
    remove all intersections as highlighted in green and orange.
    The local algorithm needs $53$ extension nodes
    (total tree energy 4438) while the global algorithm
    needs only $11$ extension nodes (total tree energy 4352).
    The intersecting trees have total energy 4334.}
  \label{fig:intersection_comparison}
\end{figure*}

In the second example, we consider the two intersecting trees
of \cref{fig:intersecting_trees}.
Each tree consists of $1,000$ nodes and the combined tree energy is $4,334$.
Without additional constraints, the trees intersect a total of $31$ times.
The result of our new global algorithm is shown in
\cref{fig:intersection_comparison} (right side)
against the local algorithm, proposed in \cite{guy20193d} (left side).
While both approaches remove all intersections,
the local one produces a large number of extension nodes
and an ``unnaturally'' high curvature of larger vessels.
In comparison, our algorithm avoids most intersections
by simply moving the branching nodes of the red tree behind the blue tree.
Thus, the cost increases only by around $0.4\%$
(using $11$ extension nodes)
compared to the local approach with a $2.4\%$ cost increase
(using $53$ extension nodes).

\clearpage
\subsection{Brain: Left frontal gyrus}
\begin{figure*}[ht]
  \centering
  \includegraphics[width=1.0\textwidth]{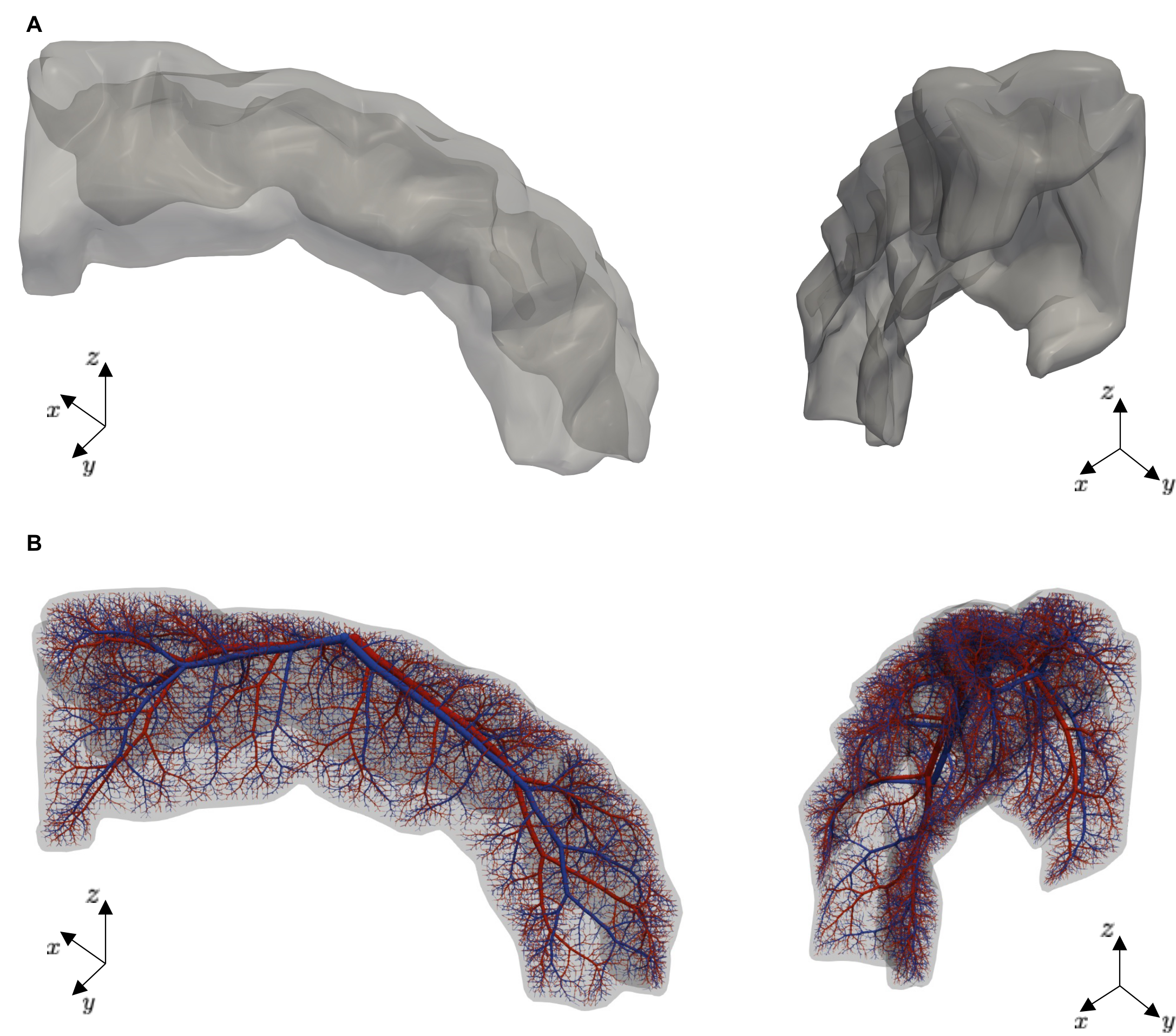}
  \caption{Generation of supplying and draining vascular trees
    inside a replica of the anatomic brain part of the left gyroid.
    \textbf{(a)} Front and side view of the left gyroid,
    highlighting the nonconvex features of the organ part.
    \textbf{(b)} Resulting vascular trees with $N\term = 50{,}000$ terminal nodes.
    Our new algorithm keeps both trees
    inside the gyroid without penetration and
    without introducing (unphysical) extensions.}
  \label{fig:gyrus}
\end{figure*}

To test how our framework handles multiple trees inside a domain with severe nonconvex features,
we choose the left frontal gyrus of the human brain.
Its geometric definition is based on a
human brain reconstruction from the BodyParts3D database \cite{bodyparts}.
We generated one inflow and one outflow tree
with 50,000 terminal vessels each.
The gyrus volume and the final generated tree are shown in \cref{fig:gyrus}.
Both trees are constrained inside the nonconvex domain
and do not intersect.
Moreover, no unnatural branches are introduced
as in local approaches \cite{guy20193d}.
Both generated trees reproduce several architectural features
observed in experimental studies
\cite{duvernoy1981cortical,lauwers2008morphometry, cassot2010branching}.
Firstly, the root quickly branches into two vessels,
which supply the left and right sides of the gyrus.
Furthermore, because the root nodes of both trees are close to each other,
the largest vessels run parallel to each other.

\subsection{Liver: Full-scale hepatic trees}
As a final example, we consider the hepatic vascular systems of the human liver.
In comparison to the brain, the liver has an almost convex shape
(unless we exlude the gallbladder).
However, in contrast to other organs, the liver has two supplying trees.
The first one is supplied through the \emph{hepatic artery} (HA) from the heart,
and the second one is supplied through the \emph{portal vein} (PV)
from the digestive tract.
The blood leaves the liver through a single draining tree
into the \emph{hepatic veins} (HV) leading into the vena cava inferior (VCI).
The synthetic generation of the hepatic vasculature
is based on the perfusion volume of the experimentally investigated trees
from Debbaut et al.\ \cite{debbaut2014analyzing}.
The physiological parameters are taken from
Kretowski et al.\ \cite{kretowski2003physiologically};
see \cref{tab:PV_parameters_2}.
We generate the vascular trees with $N\term = 100{,}000$ (matched) terminal nodes.
\begin{table*}[ht]
  \centering
  \caption{Model parameters used to generate
    the synthetic vascular trees of PV, HV and HA}
  \label{tab:PV_parameters_2}
  \begin{tabular}{*3{l@{\qquad}}l}
    \toprule
    Parameter & Description & Units & Value (PV, HV, HA) \\
    \midrule
    $\Omega$ & Perfusion volume & \si{\mm^3} & 1500 \\
    $\Tree_k$ & Initial tree & --- & Root and vessels\\
    & & & of depth $k \le 2$ \\
    $Q\perf$ & Root flow & \si{\cubic\mm\per\second} & (16.5, 20.0, 3.5) \\
    $N\term$ & Number of terminal nodes & 1 & 100,000 \\
    $m_b$ & Metabolic demand factor of blood
    & \si{\uW\per\cubic\mm} & (0.4, 0.2, 0.65) \\
    $\eta_p$ & Blood viscosity & \si{cP} & 3.6 \\
    \bottomrule
  \end{tabular}
\end{table*}

\begin{figure*}
  \centering
  \includegraphics[width=1.0\textwidth]{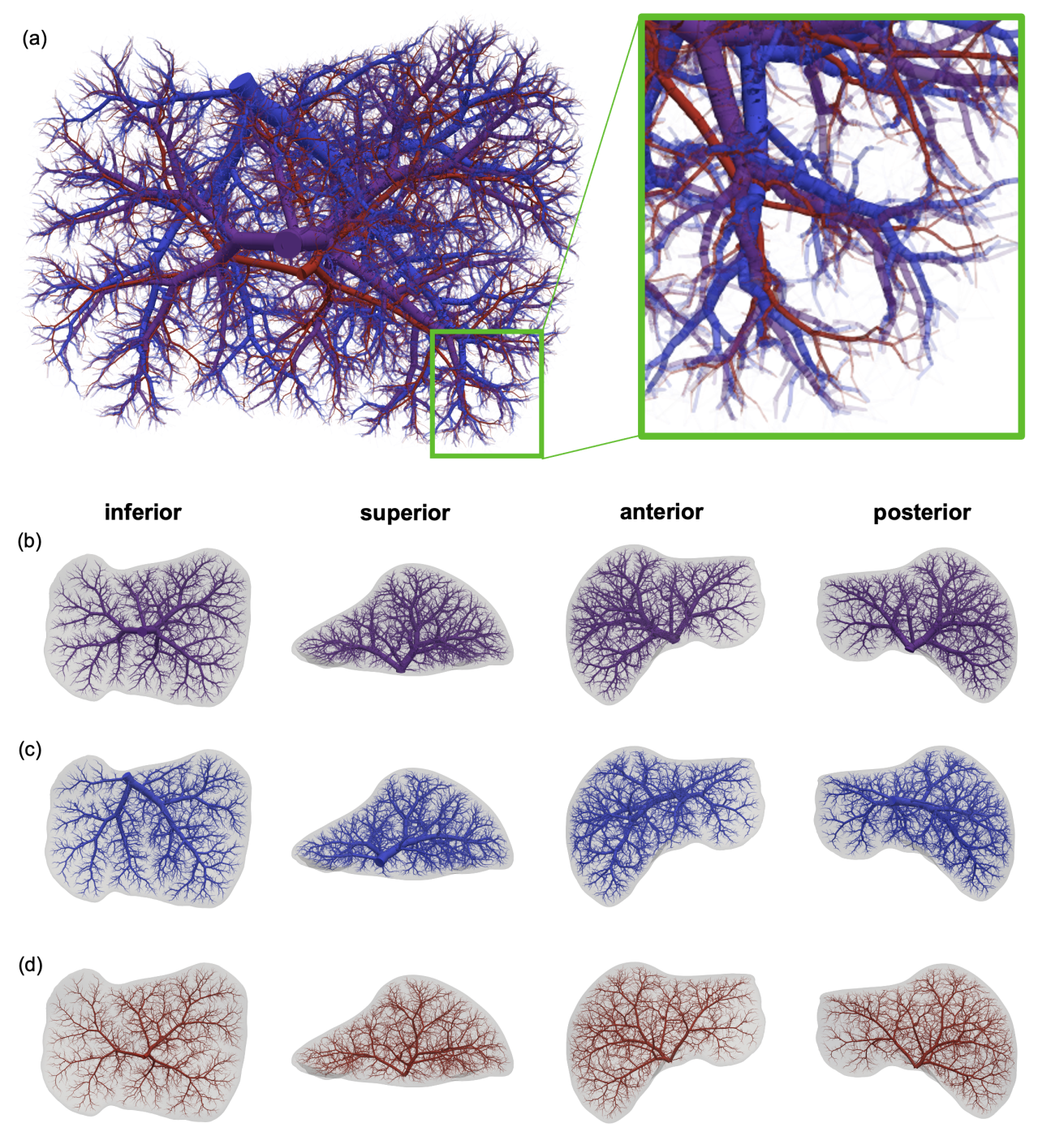}
  \caption{Synthetic hepatic vasculature (PV, HV, PA)
    generated with our optimization framework.
    Each tree consists of 100,000 terminal vessels.
    \textbf{(a)} Complete vasculature from inferior
    view with two zoom levels highlighting the different scales.
    \textbf{(b)--(d)} Individual trees inside perfusion volume (grey)
    shown from different views:
    \textbf{(b)} PV, \textbf{(c)} HV and \textbf{(d)} HA.}
  \label{fig:synthetic_liver_results}
\end{figure*}

The final hepatic vasculature is depicted in \cref{fig:synthetic_liver_results},
showing both the coupled vasculature, \cref{fig:synthetic_liver_results} (A),
and the individual trees from multiple viewpoints,
\cref{fig:synthetic_liver_results} (B--E).
We can recover multiple topological features of real hepatic vasculature trees,
most notably, the parallel development of PV and PA vessels
after the first two generations and the tendency of the smaller vessels (mainly from PA) to wrap around the larger vessels.

\section{Conclusion}
In this paper, we extended our previous framework
to generate multiple trees inside general nonconvex perfusion volumes.
We approximate regions that need to be excluded
by a set of medial balls
and introduce distance constraints to neighboring tree nodes.
Similar distance criteria are introduced
for neighboring nodes of different trees.
We include both constraint types in our global geometry optimization.
In contrast to local approaches,
we accommodate these constraints
in the global tree geometry.
This results in more natural vessel curvature and, subsequently,
reduced metabolic cost.

We tested our approach for the vasculature inside the left frontal gyroid
of the human brain and the vasculature inside the human liver.
In both cases, we can reproduce key physiological features
such as the parallel vessel paths of the larger vessels
and the tortuous shapes of smaller vessels.

While our NLP extension improves the comparison to in-vivo data, our underlying assumptions limit applicability to the pre-capillary level.
On the capillary level, the structure transmutes from pure branching to a meshed network.
This is outside the scope of our topology modeling.

We expect that our framework could be helpful in several applications.
Firstly, synthetic trees generated by this framework
could help to improve the interpretation of medical images by, e.g.,
artificially increasing the density of the initial segmented tree
to the desired pre-capillary level.
Such dense trees could improve the functional assessment of organs
such as the liver \cite{vollmar2009hepatic}.
One specific case is the understanding of why vessels become tortuous.
We can reproduce several observed shapes purely based on intersection avoidance.
Another application includes the field of tissue-engineered products \cite{salg2022vascularization}.
Here, tissue is bio-printed and requires
(ideally optimal) vascular trees to support its cells.
Our framework would allow us to construct supplying and draining trees under different design goals and constraints.

\section*{Acknowledgment}
The results presented in this paper were obtained as part of the ERC Starting Grant project ``ImageToSim'' that has received funding from the European Research Council (ERC) under the European Union’s Horizon 2020 research and innovation programme (Grant agreement No.~759001).
The authors gratefully acknowledge this support.

\bibliographystyle{ieeetr}
\bibliography{reference}

\end{document}